 \documentclass[12pt,preprint]{aastex}

\slugcomment{Not to appear in Nonlearned J., 45.}

\shorttitle{Characterizing the Purple Earth}
\shortauthors{Sanrom\'a & Pall\'e & XXXX}

\usepackage[normalem]{ulem}
\usepackage[all]{xy}
\usepackage{epsfig}
\usepackage{textcomp}
\usepackage{color}

\def\deg{\hbox{$^{\circ}$}}

\begin{document}




\title{Characterizing the purple Earth: Modelling the globally-integrated spectral 
variability of the Archean Earth}







\author{E. Sanrom\'a\altaffilmark{1,2}, E. Pall\'e\altaffilmark{1,2}, 
M. N. Parenteau\altaffilmark{3,4}, N. Y. Kiang\altaffilmark{5}, A. M. Guti\'errez-Navarro\altaffilmark{6}, 
R. L\'opez\altaffilmark{1,2} and P. Monta\~n\'es-Rodr\'iguez\altaffilmark{1,2}}

\affil{Instituto de Astrof\'isica de Canarias (IAC), V\'ia L\'actea s/n 38200, La Laguna, Spain}

\email{mesr@iac.es}





\altaffiltext{2}{Departamento de Astrof\'isica, Universidad de La Laguna, Spain}
\altaffiltext{3}{NASA Ames Research Center, Exobiology Branch, Mountain View, California 94035, USA}
\altaffiltext{4}{SETI Institute, Mountain View, California 94035, USA}
\altaffiltext{5}{NASA Goddard Institute for Space Studies, New York, NY 10025, USA}
\altaffiltext{6}{Department of Microbiology, Faculty of Biology, University of La Laguna, Spain}






\begin{abstract}


The ongoing searches for exoplanetary systems have revealed a wealth of planets with diverse 
physical properties. Planets even smaller than the Earth have already been detected, and the 
efforts of future  missions are placed on the discovery, and perhaps characterization, of small 
rocky exoplanets within the habitable zone of their stars. Clearly what we know about our planet 
will be our guideline for the characterization of such planets. But the Earth has been inhabited 
for at least 3.8 Ga, and its appearance has changed with time. Here, we have studied the Earth 
during the Archean eon, 3.0 Ga ago. At that time one of the more widespread life forms on the 
planet were purple bacteria. These bacteria are photosynthetic microorganisms and can inhabit 
both aquatic and terrestrial  environments. Here, we used a radiative transfer model to simulate 
the visible and near-IR radiation reflected by our planet, taking into account several scenarios 
regarding the possible distribution of purple bacteria over continents and oceans. We find that 
purple bacteria have a reflectance spectrum which has a strong reflectivity increase, similar to 
the red edge of leafy plants, although shifted redwards. This feature produces a detectable signal 
in the disk-averaged spectra of our planet, depending on cloud amount and on purple bacteria 
concentration/distribution. We conclude that by using  multi-color photometric observations, 
it is possible to distinguish between an Archean Earth in which purple bacteria inhabit vast 
extensions of the planet, and a present-day Earth with  continents covered by deserts, vegetation 
or microbial mats.

\end{abstract}






\section{INTRODUCTION}

Since the discovery of the first exoplanet orbiting a main sequence star in 1995 \citep{May95}, nearly 950 
extrasolar planets have been detected, and more than 2000 potential planet candidates from the Kepler 
mission are waiting to be confirmed \citep{Bat13}. In the last few years, we have been able to discover 
several planets in the super-Earth mass range (e.g. \citealt{Udr07b,Cha09,Pep11,Bor12}), some of them 
lying within, or close to, the habitable zone of their stars (e.g. \citealt{Bor12,Bar13,Ang13}). 
Even some Earth and Moon-sized planets have been recently announced 
\citep{Fre12,Mui12,Gil13,Bor13}, and this number is expected to increase in the future. In fact, early 
statistics have pointed out that around 62\% of the Milky Way's stars might host a super-Earth \citep{Cas12}, 
while studies from NASA's Kepler mission indicate that about 16.5\% of stars have at least one Earth-size 
planet with orbital periods up to 85 days \citep{Fre13}. To be prepared for the characterization of 
future discovered exoearths, first we must take a look to our own solar system and its planets.

Without a doubt, the possibility of finding life will drive the characterization of rocky exoplanets over the coming decades. 
Earth is the only planet where life is known to exist; thus observations of our planet will be a key instrument for 
characterization and the search for life elsewhere.  However, even if we discovered a second Earth, 
it is very unlikely that it would present a stage of evolution similar to the present-day Earth. 
The Earth has been far from static since its formation about 4.5 Ga ago. On the contrary, during 
this time, it has 
undergone multiple changes in its atmospheric composition, 
its temperature structure, its continental distribution, and even changes in the forms of life that 
inhabit it. All these changes have affected the global properties of Earth as seen from an astronomical 
distance. Thus, it is of interest not only to characterize the observables of the Earth as it is today, 
but also at different epochs \citep{Kal07,San12}.

Aiming at determining how Earth would look like to a hypothetical distant observer, several studies have been 
carried out over the last years. Earthshine observations have been one of the observational approaches used 
for this purpose, providing a tool to study the spectrum of Earth in the visible 
(e.g. \citealt{Goo01,Woo02,Qiu03,Pal03,Pal04}), and also in the near-infrared 
(\citealt{Tur06,Pal09}) and in the near-UV (\citealt{Ham06}). 
\citet{Ste12} studied the use of the linear polarization 
content of the earthshine to detect clouds and biosignatures.

Another possible approach is through analysis of Earth's observations obtained from remote-sensing 
platforms (e.g., \citealt{Tin06a,Cow11,Rob11,Fuj13}). \citet{Cow09} performed principal components analysis 
in order to determine if it was possible to identify surface features such as oceans and continents 
from the EPOXI data. They were able to reconstruct a longitudinally averaged map of the Earth' surface. 
\citet{Cro11} were able to categorize Earth among the planets of the solar system by using 
visible colors.

Also using EPOXI data, \citet{Kaw10,Kaw11} and \citet{Fuj12} 
proposed an inversion technique which allowed them to sketch two-dimensional planetary albedo maps 
from annual variations of the disk-integrated scattered light.

Some authors have attempted to detect the vegetation red edge through earthshine measurements 
(\citealt{Arn02,Woo02,Sea05,Mon06,Ham06}), and also using simulations (\citealt{Tin06a,Tin06b,Mon06}). 
The red edge is characterized by strong absorption in the visible part of the spectrum 
due to the presence of chlorophyll, which contrasts with a sharp increase in reflectance in the NIR 
due to scattering from the refractive index difference between cell walls and the surrounding media. 
This particular 
signature of vegetation has been proposed as a possible biomarker in Earth-like planets 
(e.g., \citealt{Sea05,Mon06,Kia07a}). The possibility of detecting hypothetical alien vegetation 
on terrestrial planets has also been studied. \citealt{Tin06c} explored the detectability of 
exovegetation in a planet orbiting an M star, on which vegetation photosynthetic pigments might 
show a shifted red edge signature. \citealt{Kia07b} conjectured further about rules for pigments 
adaptations to other stellar types.

In this paper we concentrate on the Archean eon (3.8-2.5 Ga ago), particularly on the Earth at 3.0 Ga ago  
when the Sun was about 80\% as bright as it is today \citep{Gou81,Bah01}, and the atmospheric composition of 
our planet was completely different to that of present day. At this time, the Earth's atmosphere was likely 
dominated by $N_{2}$, $CO_{2}$, and water vapor (e.g., \citealt{Wal77,Pin80,Kas93,Kas98}), with little or no free 
oxygen. Methane might have also been present as well, helping in the compensation for the reduced solar 
luminosity (e.g., \citealt{Kie87,Pav00,Haq08}).

While controversial, the first evidence of life is at 3.8 Ga in isotopically light graphite inclusions 
in apatite from Greenland \citep{Moj96}, and most likely it was non-photosynthetic, although 
this is still a subject of debate. 
The earliest photosynthetic life was probably anoxygenic bacteria like 
purple bacteria \citep{Xio00,Ols06}, utilizing reductants such as H$_{2}$ or H$_{2}$S instead of water. 
The Archean biosphere has been proposed to be a mix of anoxygenic phototrophs and chemotrophs 
such as sulfate-reducing bacteria, methanogens, and other anaerobes \citep{Kha05}. The former 
perform photosynthesis requiring a band gap energy smaller than that needed to split water, 
such that the photosynthetically active radiation relevant for anoxygenic photosynthetic bacteria 
can extend into the near-infrared to as long as $\sim$1025 nm \citep{Sch03}. Thus, their color is 
distinctly different from that of land plants that dominate the Earth today.



Because directly imaged extrasolar planets are unlikely to be spatially-resolved, we will have all planet's 
information collapsed in a single point of light. Thus, disk-averaged views of Earth are one of 
the best way to understand what kind of information one can expect from such type of observations of an Earth analogue. 
In this paper, we present disk-integrated spectra of the ancient Earth, with the aim of discerning 
the effect that a different composition of the atmosphere, and the presence of purple bacteria in different 
land/ocean configurations, might have had in the way our planet looked from afar.


\section{MODEL DESCRIPTION}

For our calculations, we make use of a line-by-line radiative transfer algorithm, based on the 
DISORT\footnote{ftp://climate1.gsfc.nasa.gov} (Discrete Ordinates Radiative Transfer Program 
for a Multi-Layered Plane-Parallel Medium) code (\citealt{Sta88}), in order to derive disk-integrated 
spectra of the early Earth. This radiative transfer model (RTM) utilizes spectral albedo of different 
surface types, profiles of atmospheric composition and temperature, 
cloudiness information, and viewing and illumination angles as input data for the calculations. 
Only a single angle of incidence and ten angles of reflection can be used for each model run. 
The RTM used here is basically an extension of 
the RTM for transits described in \citet{Gar11} and \citet{Gar12} to a viewing geometry for which the light 
reaching the observer has been reflected at the planet. With this RTM, we 
have generated a database of about 160 one-dimensional synthetic spectra that cover a wide range of 
illumination and viewing angles, and different surface and cloud types. 
Each spectrum has been calculated at 
very high spectral resolution, with at least three points per Doppler width, although, we have degraded them to a 
lower resolution (R$=10,000$) for storage purposes. Once the one-dimensional spectral library was constructed, 
we calculated the disk-averaged irradiance of the early 
Earth given a particular viewing/illumination geometry, and a map of surface properties, as described in Section 
2 of \citet{San13}.\

It is worth noting that, unlike in \citet{San13} where clouds are prescribed through a semi-empirical 
model, for the Archean there is a complete lack of reliable information on cloudiness behavior. 
Thus we have assumed that the same cloud frequency occurs at each grid point, with this cloud amount 
an input parameter in our model.


\subsection{Atmospheric Properties}

Temperature and atmospheric composition profiles for the Archean were calculated by R. Ramirez (private communication). 
A 1-D radiative-convective climate model, first developed by \citet{Kas84} and  
recently substantially updated by \citet{Kop13} and \citet{Ram13}, was used to calculate the 
atmospheric properties.

\begin{figure*}
   \begin{center}
   \xymatrix@=0.5cm{
   \includegraphics[width=0.49\textwidth]{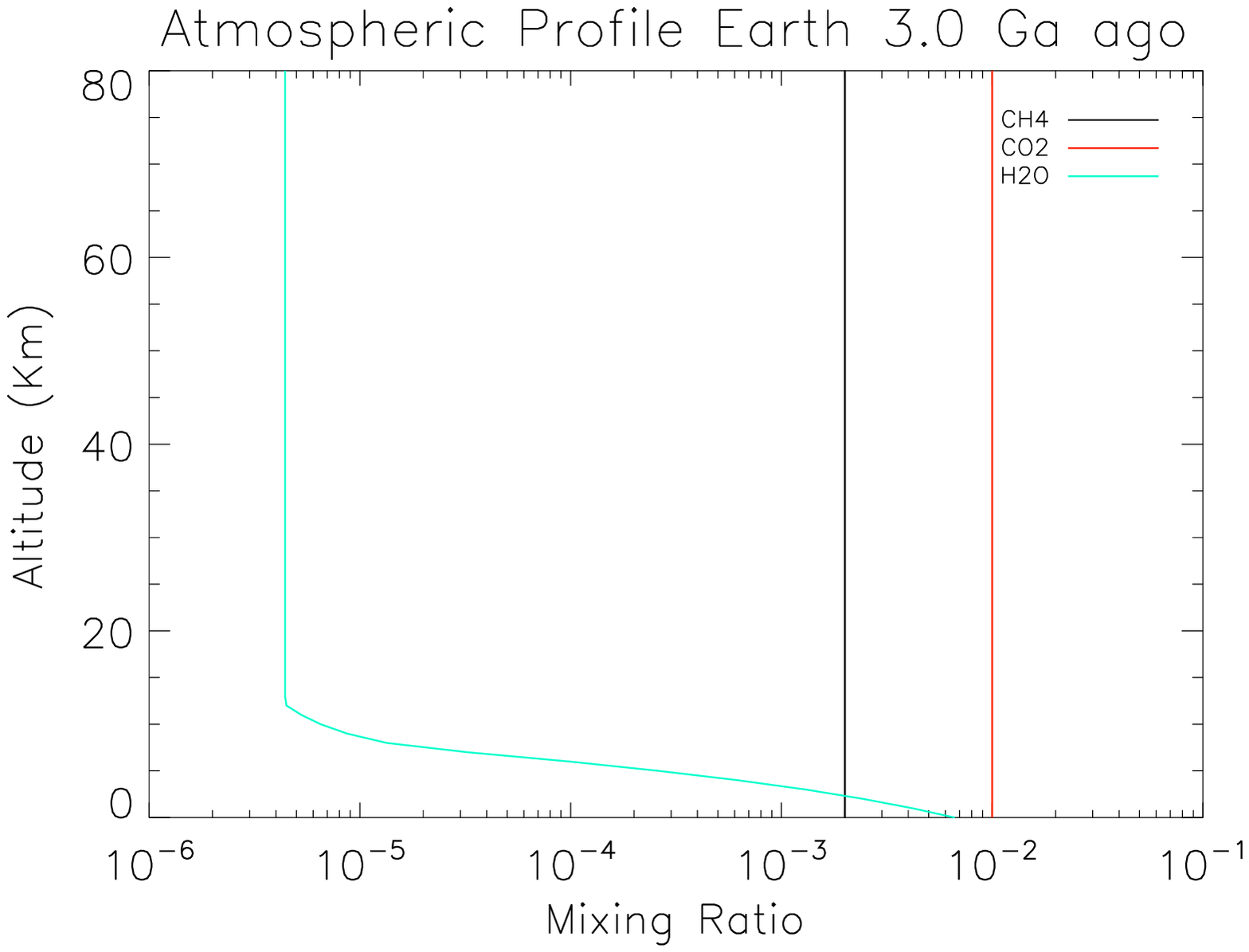} & 
   \includegraphics[width=0.49\textwidth]{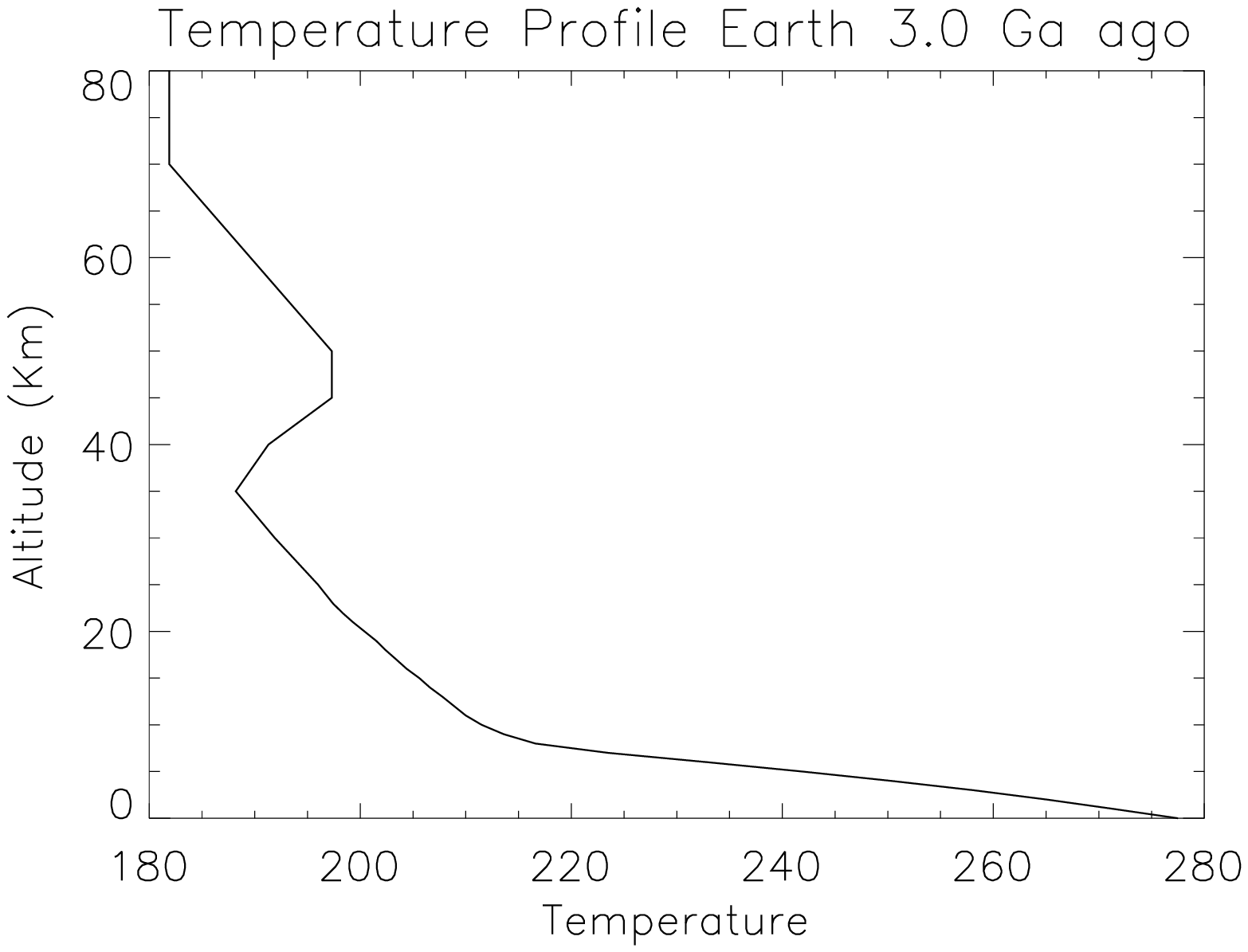}}

   \caption{Atmospheric composition and temperature profiles of the Earth 3.0 Ga ago.}
    \label{fig.early_earth_profiles}%
    \end{center}
    \end{figure*}

These atmospheric profiles consist of 1\% CO$_{2}$, 
0.2\% CH$_{4}$, according to \citet{Kal07}, being the remaining gas N$_{2}$. For the relative humidity, 
a Manabe-Wetherald profile was used \citep{Man67}. For the calculation of these profiles the Sun was assumed 
to have $\sim$79\% of its present-day luminosity, as we aimed to simulate 
the Earth 3.0 Ga ago. The temperature 
and mixing ratios profiles of these species are shown in Figure \ref{fig.early_earth_profiles}.


In our model, we divided the atmosphere into 33 uneven layers, which go from the boundary layer to 
100 Km height, with the spacing between layers of 1 km near the bottom of the atmosphere, 
and 5 km or more above 25 km height. As the original atmospheric profiles were prescribed 
in layers up to 70 Km, we assumed the same constant values between 70 and 100 Km.



\subsection{Surface Properties}

\begin{figure*}
   \centering
   \includegraphics[width=0.99\textwidth]{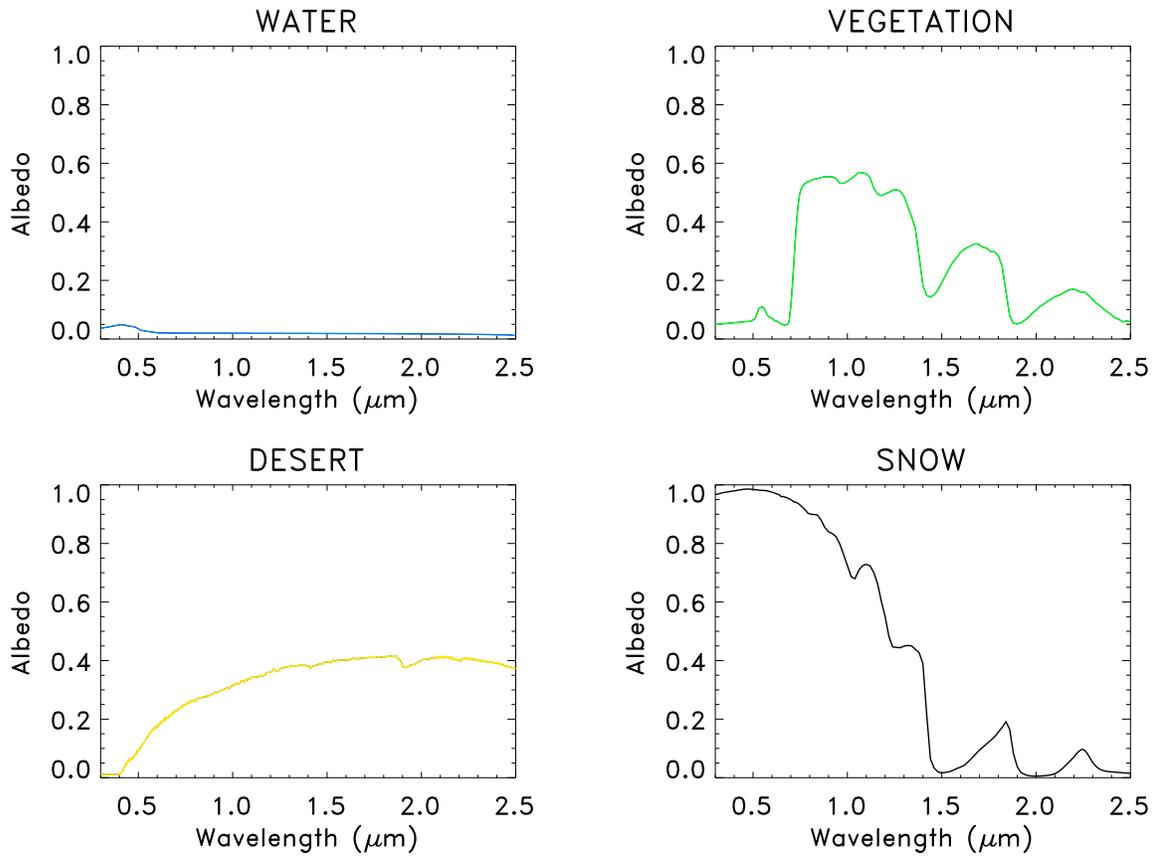}
   \caption{Spectral reflectance of water, vegetation, desert, and snow. Data taken 
   from the ASTER Spectral Library and the USGS Digital 
Spectral Library.}                                                                                              
              \label{fig.albedos}%
    \end{figure*}

To perform the disk-averaged spectra of the ancient Earth, we have considered four different planetary 
surfaces: water, desert, water with purple bacteria in suspension, and purple bacteria in microbial mats. 
There is some discussion in the literature whether purple non-sulphur bacteria could have colonized 
extended areas of soil, and whether such a signal would be remotely detectable. Here we have assumed 
as the most likely scenario, that these microbial mats are located in marine intertidal environments. 
The wavelength-dependent surface reflectivity of the two first surface types were derived from the ASTER 
Spectral Library\footnote{http://speclib.jpl.nasa.gov} and the USGS Digital Spectral 
Library\footnote{http://speclab.cr.usgs.gov/spectral-lib.html}. Figure \ref{fig.albedos} 
shows the spectral albedo of these surface types. The wavelength-dependent albedo of 
surfaces involving purple bacteria 
were obtained as described in Section \ref{sec.bac} (Figure \ref{fig.albedo.purple}). \

\subsection{Surface Distribution}

The movement of Earth's tectonic plates has caused the formation and the break up of continents over 
Earth's history, including the formation of supercontinents. The orientation of Earth's earliest 
continents is still unknown, although it is believed that the fraction of the surface covered 
by continents during the Archean was smaller than present day (e.g. \citealt{Goo81,Bel10,Dhu12}). 
Earth might have been almost entirely covered by water with some small continents. Hence, due to the 
complete lack of information about the continental distribution of the Earth 3.0 Ga ago, we decided to use 
that of the Earth during the Late Cambrian (500Ma ago, see Figure \ref{fig.mapas}) and at present day, as 
two possible characteristics examples. 
The continental distribution 
of these two epochs has been taken from 
Ron Blakey's website\footnote{http://jan.ucc.nau.edu}, where these surface maps are available online. 
The Earth geologic information has been regridded into the 64x32 pixel grid used by our model.

We have also defined the coastal areas of these maps. These coastal zones were determined 
as those land grid cells that have adjacent ocean grid cells and those ocean cells that have adjacent 
land grid cells.


\subsection{Cloud Optical Properties}

For these simulations we have taken into account three different cloud layers: low (1000-680 mb), mid (680-440 mb) 
and high cloud (440-30 mb). The optical properties of each cloud type, wavelength-dependent scattering and 
absorption coefficients, and the asymmetry parameter, were taken from the Optical Properties of Aerosols and 
Clouds (OPAC) data base (\citealt{Hes98}).
We have considered physical cloud thicknesses of 1 km, and we have assumed that the scattering phase function
is described by the Henyey-Greenstein equation inside clouds, and by the Rayleigh scattering function outside them.\ 

The atmospheric composition of our planet has not changed drastically in the last 500 Ma. On average, 
the same atmospheric composition and mean averaged temperature have existed during this period 
\citep{Har78,Kas02}. Hence, in \citet{San13} we used a semi-empirical model of clouds in order 
to reconstruct the possible cloud distribution of the Earth in the past, up to 500 Ma ago. 
However, this does not hold true for longer period of time, and such cloudiness reconstruction 
is not a valid approximation for the Archean. Thus, we prescribe the mean cloud frequency of the Archean as an 
input parameter, and explore how different levels of cloudiness may have affected the way our 
planet looked from afar, and in particular, the effect that could have had in the possible 
detection of biomarkers. 
Note that detailed cloud structures, tied to ocean currents or continents, are missing in or simulations. 
However, for globally-integrated spectra like the ones investigated here, this limitation is small.

%

\section{PURPLE BACTERIA'S REFLECTANCE SPECTRA }
\label{sec.bac}



To obtain the reflectance spectra of purple bacteria, we used pure cultures of 
\textit{Rhodobacter sphaeroides} ATCC 49419, a purple non-sulfur bacterium growing as a suspension of cells in liquid 
media. This type of phototroph exhibits diverse metabolic abilities, allowing survival in a wide 
range of dynamic environmental conditions. Purple non-sulfurs can grow aerobically in
the dark as a chemoheterotroph, and anaerobically in the light using hydrogen and 
organic compounds as electron donors for photosynthesis.

We made two different measurements to retrieve the reflectance spectrum of these bacteria. In the first one, 
the reflectance of a liquid pure culture of purple bacteria was measured using a UV/VIS/NIR 
spectrophotometer (VARIAN, CARY 5E). The measurements were taken from 0.3 to 2.5 $\mu$$m$. 

In the second experiment, we used a LiCor LI1800 spectroradiometer with a remote cosine receptor 
that was positioned 5 cm above the culture. These measurements cover the 0.35-1.1 $\mu$$m$ spectral range. This culture of purple bacteria was in a petri dish sitting on top of a piece of white paper. In order to calculate the spectral albedo, the spectral irradiance of the Sun was also measured.

\begin{figure*}
   \centering
   \includegraphics[width=0.99\textwidth]{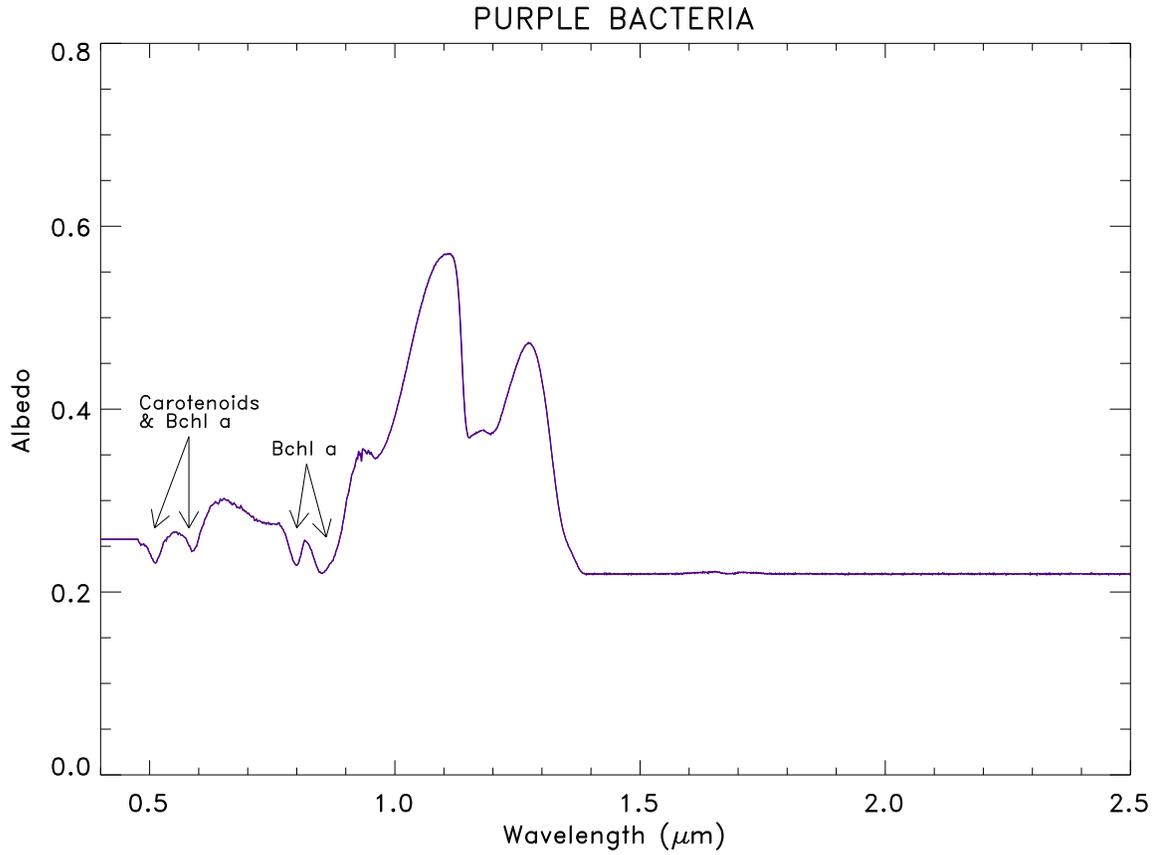}

   \caption{Wavelength-dependent albedo obtained for purple bacteria in the VIS-NIR spectral range where major absorption features of carotenoids and \textit{bacteriochlorophyll a} are labeled}.                                                                                              
              \label{fig.albedo.purple}%
    \end{figure*}

\begin{figure*}
   \begin{center}
   \xymatrix@=0.5cm{
   \includegraphics[width=0.49\textwidth,height=5cm]{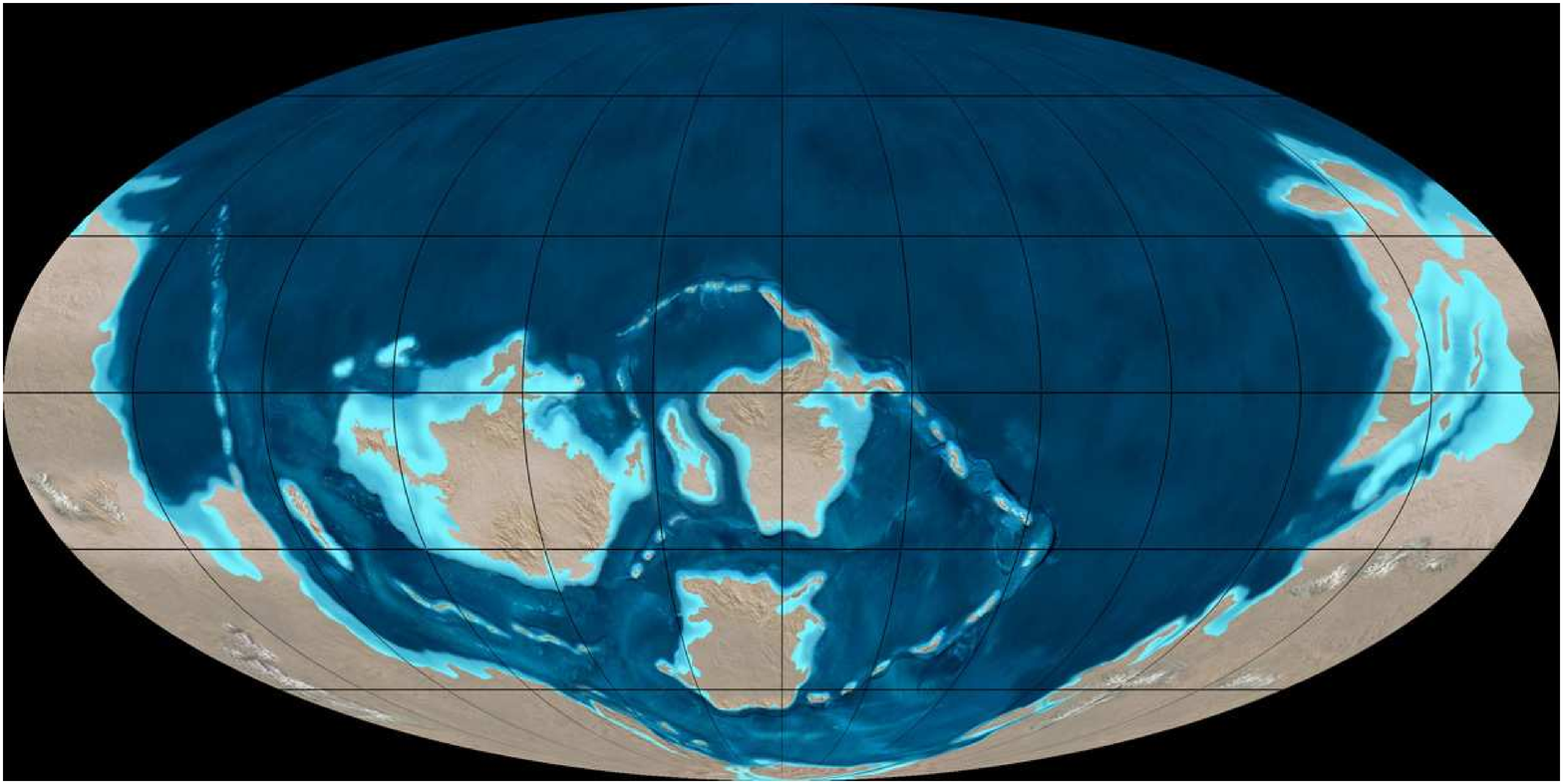} & 
   \includegraphics[width=0.49\textwidth,height=5cm,trim=30 0 1 0, clip]{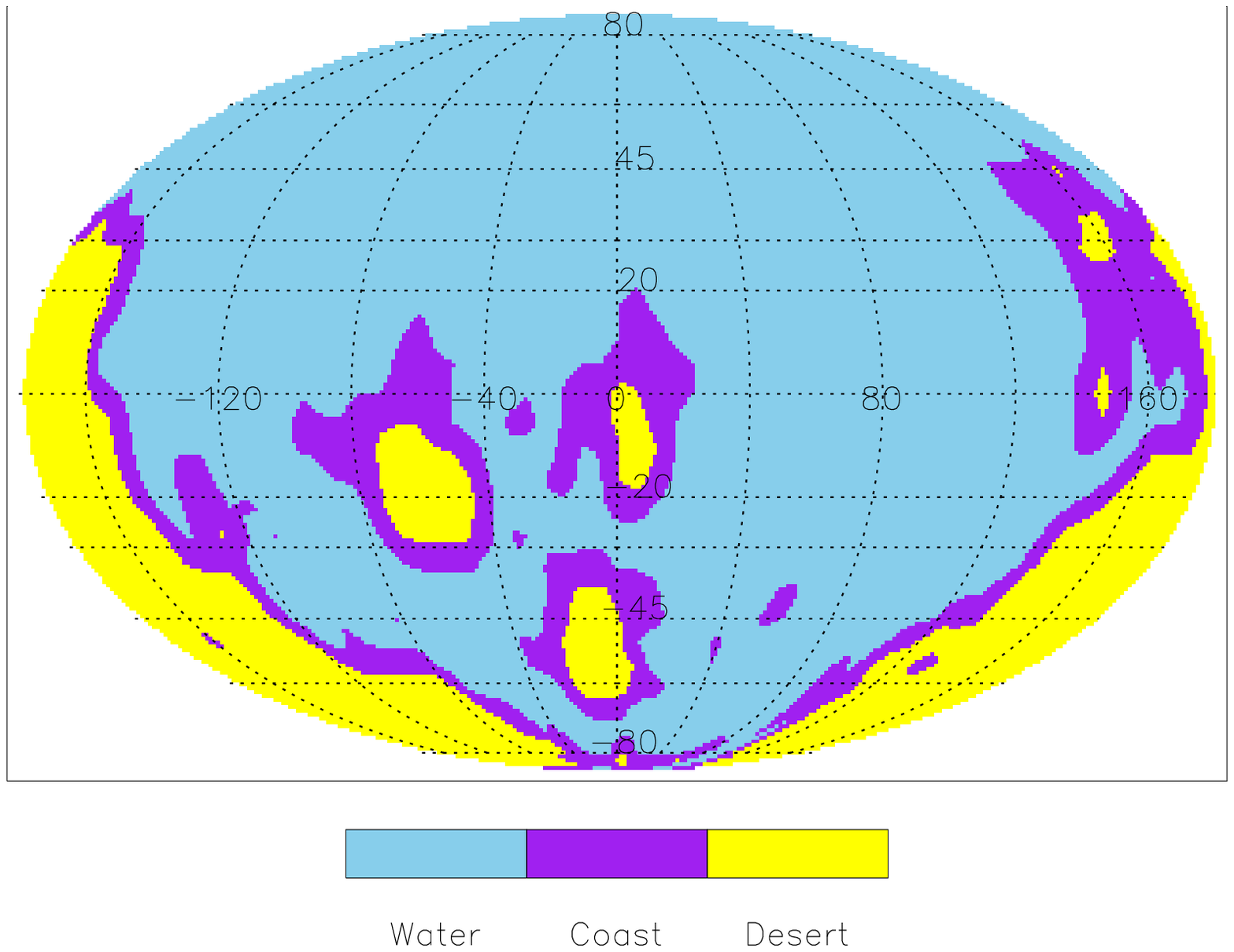}}

   \caption{Left: A map of the Earth's continental distribution during the Late 
    Cambrian (500 Ma ago). Image credit, Ron Blakey. Right: Same as in left panel 
    but here oceans, continents and coastal zones are indicated in blue, yellow and purple, respectively.
    Data are plotted with a geographical resolution of 64x32 grid cells (longitude by latitude), 
    the same that our models use. Coastal areas constitute 14\% of the total grid cells when using 
    this continental distribution and geographic resolution. 
    In our simulations the 00:00 hours UT, correspond to the subsolar point crossing the image center.}

    \label{fig.mapas}%
    \end{center}
    \end{figure*}

Both sets of retrieved reflectances spectra were merged into one, covering from the visible to the near-infrared  (Figure \ref{fig.albedo.purple}), using as a reference the second experiment, which had a higher signal to noise ratio. In order to absolutely calibrate the reflectance spectra of the bacteria, we also measured the 
reflectance spectra of a set of known leaves. The comparison of our measured leaf reflectance 
spectra with those tabulated in the ASTER library gave us a measure of our reflected flux, 
which we then applied to the reflectance spectra of the purple bacteria. 
This way we transformed the ratio scale into a reflectance scale.


Figure \ref{fig.albedo.purple} shows the reflectance spectrum of the purple non-sulfur bacterium \textit{Rhodobacter sphaeroides}. The photosynthetic pigments of these bacteria are \textit{bacteriochlorophyll a} esterified with phytol, and carotenoids of the spirilloxanthin series. Due to the combination of these two pigments, living cells of this species show absorption features at 375, 468, 493, 520-545, 589, 802, 860-875 nm \citep{Imh05}. Some of these features can be detected, and are marked, in Figure \ref{fig.albedo.purple}. Moreover, as the purple non-sulfur bacteria cultures used were red, the reflectance increase between $\sim$600 and 700 nm is due to the reddish light reflected back from the cultures' cells. The most noticeable feature of this spectrum is the sharp increase in 
reflectivity from approximately 0.9 $\mu$$m$ to 1.1 $\mu$$m$, and the equally strong decrease from 1.3 $\mu$$m$ to 1.4 $\mu$$m$. Information about the physical nature of the absorption features at $\lambda$ $>$ 1 $\mu$m is not found in the literature. Starting at 1.4 $\mu$$m$ and redwards, the spectrum does not show any measurable features, and the overall albedo value is probably that of water, made slightly more reflective due to the presence of bacteria in suspension that lower its transmissivity. The overly-featureless variability is probably due to the low sensitivity of the instrument used at these wavelengths.

The bacteria concentration in our sample culture was very high ($\sim$10$^{9}$ cells/ml), probably much more than the typical concentrations that would be found in seawater. Thus, when we modeled purple bacteria in open oceans, we used a combination of pure seawater and our bacteria(+water) spectra, weighted in varying percentages, to simulate different bacteria concentrations. Throughout the rest of the paper we refer to percentage dissolution (for example a dissolution of 10\% means 9 parts of seawater and 1 part of our culture, equivalent to concentrations of the order of $\sim$10$^{8}$ cells/ml). Note that as we used liquid cultures to measure the reflectance of purple bacteria, the effect of the transmittance of water is already included in the spectrum.



\section{THE SPECTRA OF THE EARLY EARTH}

\label{fig.500Ma_100_10_nubes}

In order to determine if it would be possible to discern the presence of life forms such as purple 
bacteria in the spectra of an extrasolar planet, we have simulated the disk-integrated spectra of the 
ancient Earth, taking into account different continental distribution, cloud coverage, and several 
abundance scenarios which go from a planet where purple bacteria have colonized both oceans and continents, 
to a planet where purple bacteria are only found in oceanic coastal areas in low concentrations.

In all the cases studied in this manuscript, both the observer and the Sun are located over the planet's equator 
in such a way that the observer is looking at a half-illuminated planetary disk, i.e., at a phase angle of 90\deg. This 
is the most relevant geometry for studying exoplanets, since the maximum angular separation of an extrasolar planet 
from its parent star along its orbit, takes place at phase 90\deg, as defined from the observer's position.

\begin{figure*}
   \begin{center}
   \xymatrix@=0.5cm{
   \includegraphics[width=0.49\textwidth]{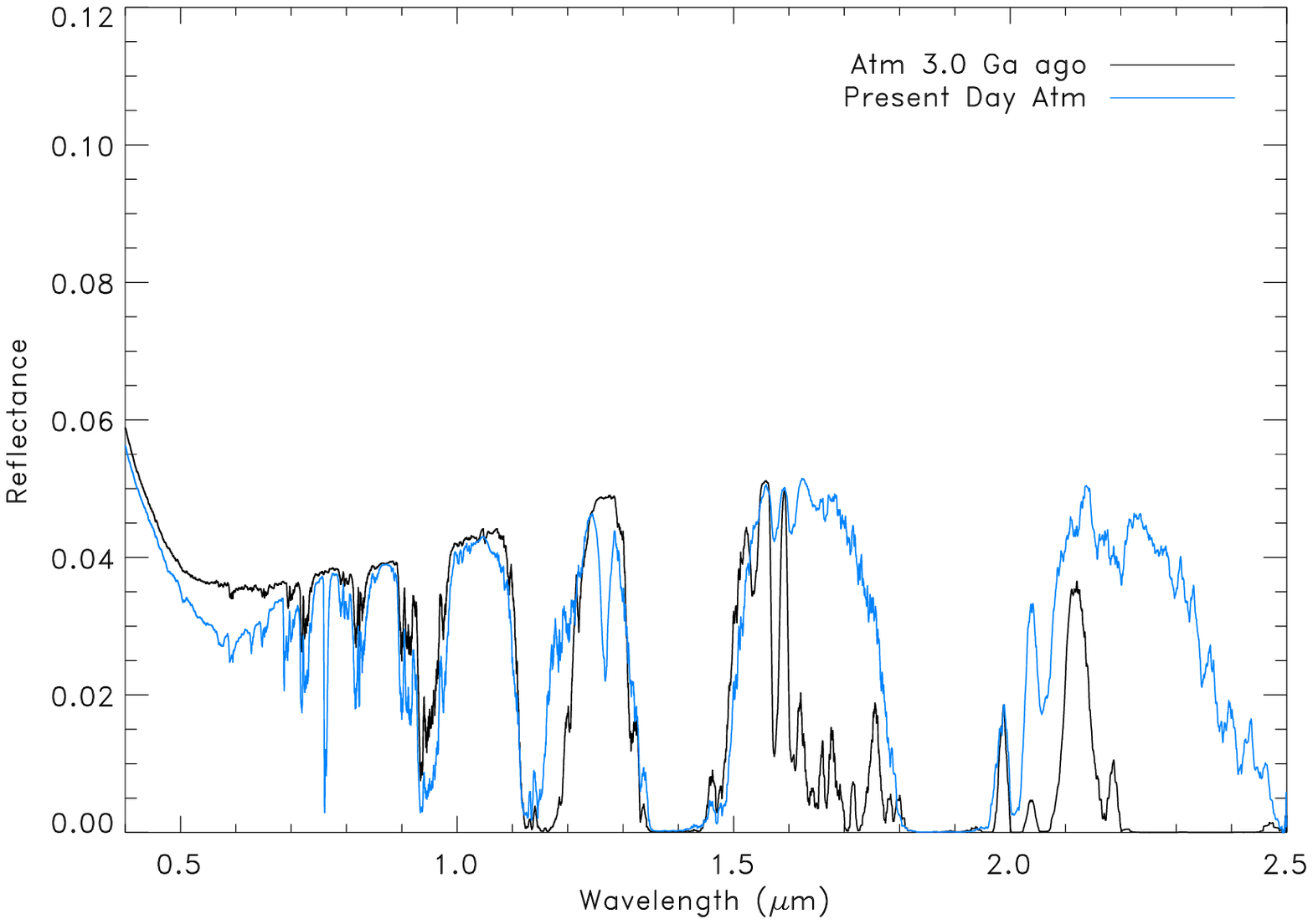} & 
   \includegraphics[width=0.49\textwidth]{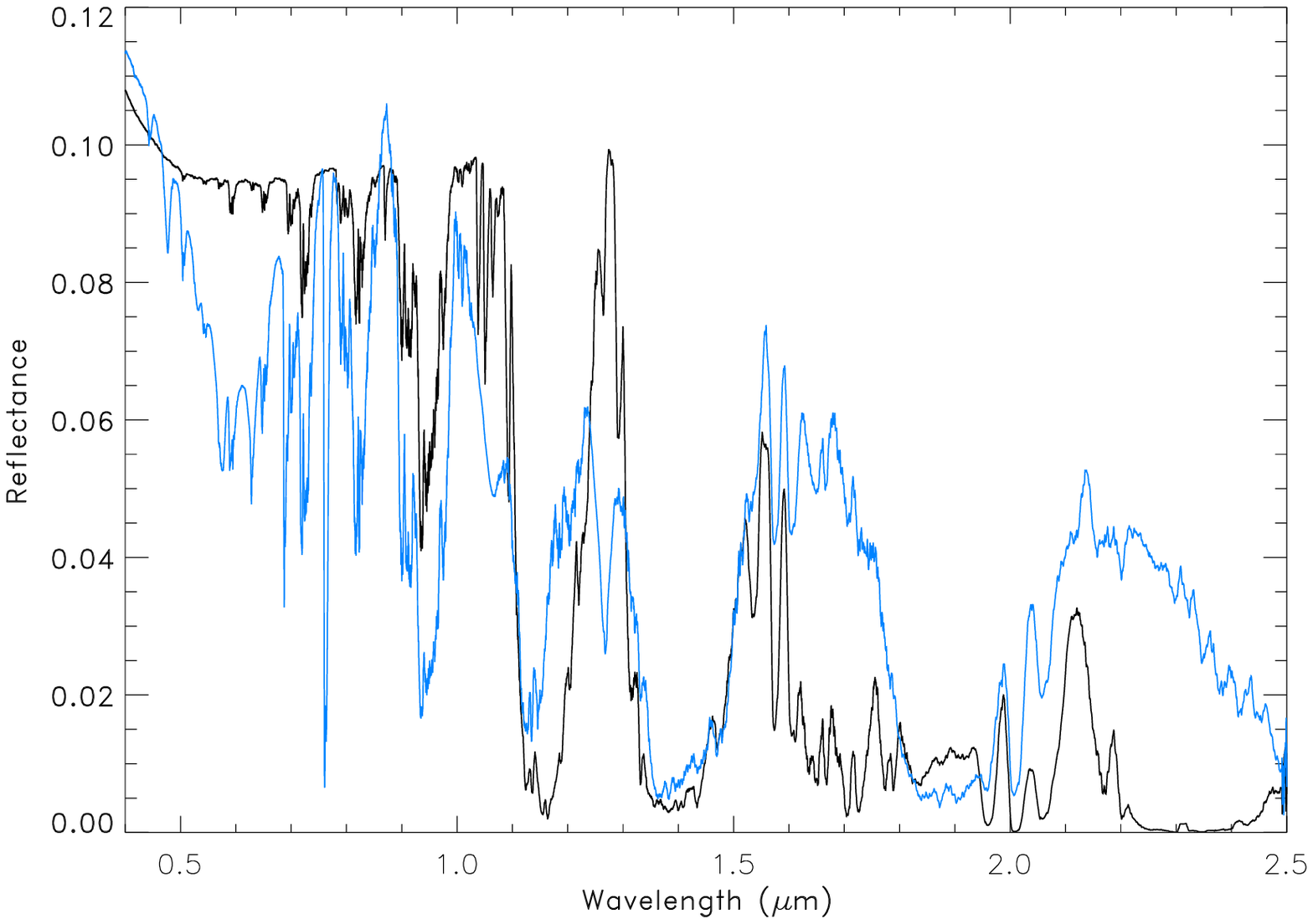}}

   \caption{Visible and near-infrared disk-averaged spectra of a planet with a continental distribution 
    of the Earth 500 Ma ago, with an atmospheric composition similar to that of the Earth 3.0 Ga ago 
    (black), and with present-day composition (blue). Here, continents are totally desert and we have 
    considered both, a cloud-free atmosphere (left), and a cloudy atmosphere (right). We have assumed 
    that clouds cover $50\%$ of the surface. The spectra have been smoothed with a 100 point 
    running mean for display purposes.}
    \label{fig.500Ma_atmosferas}%
    \end{center}
    \end{figure*}

In order to see the effect of considering different atmospheric compositions in the disk-averaged 
spectra of a lifeless planet, Figure \ref{fig.500Ma_atmosferas} compares the spectrum of a planet with 
a CO$_2$-CH$_4$ rich atmosphere and no oxygen (black), with the spectrum of a planet 
with Earth's current atmospheric composition (blue). In both cases, 
the continental distribution is that of Earth 500 Ma ago and continents are assumed to be completely 
covered by deserts. Figure \ref{fig.500Ma_atmosferas} (black lines) show strong absorption in the NIR 
part of the spectrum due to the increased levels of CO$_2$ and CH$_4$, while in the visible region, 
the most noticeable difference with Earth's current atmosphere is the lack of the absorption 
features typical of O$_2$ and O$_3$.

\subsection{The effects of clouds}

The visibility of surface inhomogeneities, such as continents or surface types, on a planet 
is naturally very dependent on the frequency of cloud formation. 
The top panels of Figure \ref{fig.500Ma_100_10_nubes} shows synthetic disk-averaged spectra of 
the ancient Earth 
over a course of a day, one spectrum every two hours, covering the spectral range between 0.4 and 2.5 $\mu$$m$, 
for both a cloud free (left) and a cloudy atmosphere (right). Cloud cover is assumed to be 50\%. 
In the top panels, the continental distribution corresponds to that of Earth 500Ma ago, 
and continents are dry 
lands (deserts), while the coastal points closest to land are completely covered by purple bacteria, and the 
coastal points closest to ocean are a mixture of 10\% purple bacteria and 90\% of water. 
We have chosen this particular scenario since bacteria are expected to be found 
where nutrients are more abundant, like in shallow waters or coastal zones. 

It is expected that as continents come in and out of the field of view, the 
light reflected back by the Earth changes, with these changes more drastic 
for the cloud-free cases. The reflectance is higher when continents 
occupy most of the observable half-disk (at 8:00-10:00 UT, when the percentage of continental surface in the sunlit area of the planet visible from our observer's location is $\sim$50\%), than when oceans dominate the field of view (at 16:00-18:00 UT, when the percentage of continental surface is $\sim$95\%). 

The addition of clouds to the model results in a overall enhancement of the light reflected back by the planet,
and a strong decrease in the reflectance variability over a diurnal cycle. This results in a significant loss of information about the surface types lying under clouds. It is worth noting that changing cloud cover or even varying cloudiness 
distribution could significantly change the overall shape of these disk-averaged spectra.

\subsection{The effects of continental distribution}

The land-mass distribution of the Earth 500 Ma ago consisted of a large continental mass, mostly 
located in the southern hemisphere. There is also a 
further group of three large islands also in the southern hemisphere. 
On the other hand, the present-day Earth has two major continental land masses spreading over the north 
and southern hemispheres. 
In order to estimate the effect of considering different land-mass distributions, 
Figure \ref{fig.500Ma_100_10_nubes} bottom panels 
show the same spectra as the top panels, but here the input 
continental distribution corresponds to that of the present-day Earth. Although these two 
continental distributions are considerably different, the disk-integrated spectra of both cases 
are quite similar. The rotational variability is also similar in both cases, with a comparable amplitude, 
only slightly smaller for the present day Earth when clouds are considered. 

\subsection{Abundances and detectability}

Figure \ref{fig.bac_10_noCL_conCL} (left) shows disk-integrated spectra obtained for the early Earth where continents 
are completely covered by purple bacterial mats, and oceans are a mixture of water and bacteria, 90\% and 10\%, 
respectively. Both a cloud-free (red) and a 50\% cloudy case (black) are shown. The continental distribution 
used here is that of Earth 500Ma ago. 

Whether purple bacteria would have been able to colonize the continental surfaces during the Achean is still unsolved. Due to the lack of $O_{2}$ in the early Earth's atmosphere, and therefore the lack of a ozone layer, harmful radiation did probably reach the Earth's surface, and purple bacteria might have suffered from DNA damage. Studies of modern microbes, however, suggest that their photoprotective pigments, that absorb in the blue and UV (e.g., carotenoids), are sufficient to have allowed for their survival in terrestrial and shallow water environments on the early Earth \citep{Coc98}. Moreover, some purple bacteria have been shown to use reduced iron (Fe(II)) for photosynthesis, and \citet{Pie93} pointed out that the oxidized iron products of this type of photosynthesis could have provided substantial protection from UV radiation for surface-dwelling phototrophs prior to the development of an ozone shield. Thus, the presence of bacterial mats in continental areas, while not being our most likely scenario, cannot be ruled out.

In Figure \ref{fig.bac_10_noCL_conCL} (left) we only show the spectra at the time when 
the continental presence is at maximum, i.e., at 08:00 UT, when the percentage  of land over 
ocean that is illuminated and visible at the same time is about 50\%. The same is shown in 
the right panel of this Figure, but here considering that purple bacteria do not exist over 
continents, and are only found in the oceans. For comparison, Figure \ref{fig.500Ma_atmosferas} 
(black lines) shows the spectra of a planet without bacteria in either water or land.

When the amount of purple bacteria is high, 100\% over continents and 10\% diluted in the water, the 
presence of purple bacteria on the early Earth produces a strong feature, 
a steep increase in reflectance around 
1.0 $\mu$$m$ in its disk-integrated spectrum 
(Figure \ref{fig.bac_10_noCL_conCL} left). When clouds are included in the model, this signature is 
naturally significantly 
diluted. However, it is still easily detectable by simple inspection of the disk-average spectra. 

When considering a more realistic case where purple bacteria are only found in coastal areas
(Figure \ref{fig.500Ma_100_10_nubes}), the increase in reflectivity around 1.0 $\mu$$m$ due 
to these bacteria is readily seen in the cloud-free case and is still detectable in the cloudy case.

For a planet with marine bacterial life only, and bare continental surfaces, 
this spectral feature produced by purple bacteria becomes 
harder to discriminate in the spectra (Figure \ref{fig.bac_10_noCL_conCL} right), 
being practically undetectable in the cloudy case. Here we have considered a mixture of 
10\% bacteria and 90\% water. In fact, if one compares the cloudy spectrum of this case with a case where there are no 
bacteria, neither over continents nor in the water (Figure \ref{fig.500Ma_atmosferas}, black lines) 
it is impossible to distinguish 
between them by simple exploration of the spectra. 
Thus, to estimate the marine-only purple bacteria detectability, we have considered several bacteria 
concentrations in oceans: 20, 30, 40, and 50\%. Table 
\ref{tbl-1} shows the slope of the straight line that connects the averaged planetary radiance 
in the 0.745-0.770 $\mu$$m$ and 1.010-1.034 $\mu$$m$ spectral intervals. 
The data are given as a function of bacteria concentration in water for both a cloud-free and a 
cloudy atmosphere. 
The slope between these two spectral regions, free of atmospheric absorption features, is mainly 
influenced by the contribution of purple bacteria to the globally-integrated spectrum of Earth. 
Thus, this slope can be used as a measure of the strength of the purple bacteria signal, similar 
to what has been previously done with to quantify the vegetation's red edge \citep{Mon06}. 
Table \ref{tbl-1} shows how 
increasing the bacteria concentration in water from 10\% to 50\% monotically increases this slope. 
Although not shown here, 
the identification of the presence of purple bacteria by simple inspection of these spectra in 
the cloudy atmosphere case is almost impossible for bacteria concentrations in oceans lower than 30\%.

Finally, although it is a very improbable scenario, we have run a comparison test to estimate the 
purple bacteria detectability. Figure \ref{fig.bac_chl} shows disk-averaged spectra of Earth 
considering the present-day 
continental and cloud distribution, and the early Earth atmosphere. Cloud distribution was 
taken as the 1984-2006 climatology of ISCCP\footnote{http://isccp.giss.nasa.gov} cloudiness data. 
Here we have assumed that continents are completely deserts and we have used the 2012 annual mean 
ocean chlorophyll a content map from the SeaWiFS project\footnote{http://oceancolor.gsfc.nasa.gov/} 
as a proxy for the distribution of purple bacteria. 
Thus, we have 
considered that the ocean latitudinal range 90\deg \texttwelveudash 35\deg, both Nord and South, and the 
-10\deg \texttwelveudash 10\deg latitudinal range are a mixture of 90\% bacteria and 10\% of water. 
Coastal zones are also populated with purple bacteria, and the rest of the ocean is a mixture of 
10\% bacteria and 90\% water. The figure shows the spectra at 8:00 UT (when oceans dominate the field of 
view; blue lines) and at 18:00 UT (when continents dominate the field of view; black lines). As in previous 
cases in a cloud-free atmosphere purple bacteria are readily detectable. For a present-day cloud amount 
(roughly 60\%) the detectability is not so obvious, and a high signal-to-noise ratio spectra of the planet 
would be needed.


\section{Photometric Light-Curves}
\label{sec_photometric}

In the case of a terrestrial planet in the habitable zone of a G, F or K star, obtaining in the near-future 
even a low resolution spectra might be a difficult task to perform 
(e.g., \citealt{Pal11,Rug13,Hed13}). 
Photometric observations on a few filters might actually be a more realistic possibility, 
even if the data are obtained via spectroscopic observations, as is the case of 
Hubble and Spitzer observations nowadays (e.g., \citealt{Tin07,Swa08,Des11,Pon13}). 
The fact that a planet shows photometric variability along one rotation already speaks about 
the presence of inhomogeneities in its surface or atmosphere \citep{For01}. 
With sufficiently accurate time series it might be possible to distinguish the presence of 
cloudiness and continents based on this variability \citep{Pal08}. 
Moreover, the rotational photometric variability 
as a function of wavelength can reveal information about the major wide-spread composition of 
continental surfaces.

Thus, following \citet{San13}, we have convolved our modeled disk-integrated spectra against 
standard astronomical filters, both visible and near-infrared, namely B, V, R, I, z, J, H, and K. 
Figure \ref{UBVRI_500D_0cl} shows the photometric daily variations in each photometric filter of the disk-averaged 
reflected light, for a cloud-free atmosphere (red) and for a cloudy one (black). 

%
%
%

For these simulations, we have used a conservative scenario: the continental distribution assumed is that of 
Earth 500 Ma ago, continents are bare desert, coastal land areas are covered by purple bacteria mats, 
oceanic coastal areas  are a mixture of 10\% purple bacteria and 90\% of water, and open ocean areas are only water. 

The light curves shown in Figure \ref{UBVRI_500D_0cl} all have a similar shape; the peak in brightness takes place in each 
of the photometric filters when continents dominate the field of view, between 8:00 and 10:00 UT, while minimum values of reflectance occur when oceans occupy most of the view, approximately at 4:00 and at 16:00 UT.

The cloud-free case shows a larger variability in the diurnal light curves in each filter and a considerable 
rise in brightness that increases monotically redwards. A similar result is found when clouds are added to the model, 
although it shows much less variability owing to the effect of clouds on the spectra. 

Figure \ref{fig.BVRI_zJHK} (top) shows the amplitude of albedo variations of the 
two scenarios shown in Figure 
\ref{UBVRI_500D_0cl} as a function of the different photometric filters. As found before, the variability 
of the light reflected back by the planet in both the cloud-free and cloudy cases increases toward the red, 
with this increase much more dramatic in the cloud-free case than in the cloudy one. The amplitude of these 
variations goes from a few percents to around 180\% and 50\%, for a cloud-free and a cloudy atmosphere, 
respectively.

Figure \ref{fig.BVRI_zJHK} (bottom) shows the expected photometric variability for a modern Earth with different surface compositions 
(data from \citealt{San13}), together with the Archean Earth results for the cloudy case (from top panel). The different color lines show the amplitude of the albedo variability for a cloudy atmosphere with 
an atmospheric composition similar to that of present Earth, for a planet totally desert (black line), completely 
covered by vegetation (green line), by microbial mats (red and dark blue lines), and for a planet where continents are a mixture of microbial mats and deserts (yellow line), and microbial mats and vegetation (cyan line). In all cases the atmospheric composition in the models used to generate the data is that of the modern Earth. The purple line shows the albedo variability of a cloudy planet with an atmospheric composition similar to that of the early Earth (3.0 Ga ago) which continents are covered by deserts and coasts are a mixture of purple bacteria and water. 

In \citet{San13} we concluded that it would be possible to discriminate between vegetated continents, large 
extension of microbial mats and bare continental surfaces by comparing the amplitude of the albedo change, 
along the course of a day, taken in different photometric filters. Here, the purple curve suggests 
that it would also be possible to discriminate a purple Archean-like planet among the other scenarios studied in 
\citet{San13}. In contrast to the other scenarios, in the visible portion of the spectrum, the amplitude of albedo change of the Archean Earth increases monotically from the B filter to the z filter, an effect due to the combination of the lack of oxygen in the atmosphere and the reflectance spectra of the purple bacteria. When one 
moves towards the red part of the spectrum, this increase is sharper and the curve peaks at the H filter 
to decrease redwards in K. However, if one only takes into account the near-IR filters, it would be difficult 
to discriminate between a desert planet and the purple scenario.


\section{CONCLUSIONS}

In this paper we have presented disk-integrated spectra of the Earth during the Archean eon with the 
aim of studying how a different atmospheric composition and the presence of early life forms, such as 
purple bacteria, may have affected the way our planet looked from afar.

As one of the inputs of our models is the reflectance spectrum of different surface types, we 
carried out two different experiments in order to retrieve the spectral albedo of these bacteria in 
the spectral rage 0.3-2.5 $\mu$m. We found that purple bacteria 
show a reflectance spectrum with a sharp increase in reflectivity similar to the red edge of 
leafy plants, but shifted redwards.
 
In order to determine if it would be possible to detect such a biomarker, 
we have considered three different scenarios: one where purple bacteria have colonized the whole 
planet, both water and continents, other where purple bacteria are only found in oceans, and finally 
a scenario where these bacteria are found only in coastal regions. We have taken into account 
the effect of clouds in our models finding that the inclusion of clouds results in 
the increase of the reflectivity of the planet, reduces drastically the albedo variability 
over the course of a day, and makes more difficult the identification of surface types under clouds. 
Changing the continental distribution does not seem to have a high impact in the globally-averaged 
spectral variations.

We find that when the amount of purple bacteria is high, they can be readily detected in 
disk-averaged spectra, both in cloud-free and in cloudy atmospheres. While if purple bacteria are only 
found in oceans, their spectral feature becomes nearly undetectable in the cloudy case. 
When considering a more realistic 
scenario where purple bacteria are found in coasts, their presence can be 
detectable in the cloud-free case, and even in the cloudy case, although the signal is smaller.

Finally, by convolving these simulated spectra against standard astronomical filters, 
we conclude that using photometric observations in different filters might allow us to discriminate between a 
present day Earth with continental surface covered by deserts, vegetation or microbial mats, from an Archean 
Earth where purple bacteria have colonized large extensions of the planet.


\acknowledgments

\textbf{Acknowledgments}\\

We would like to thank Ant\'igona Segura and Ramses Ramirez for kindly providing us 
with the atmospheric profiles of the early Earth.









   \begin{figure*}
   \begin{center}
   \xymatrix@=0.5cm{
   \includegraphics[width=0.49\textwidth]{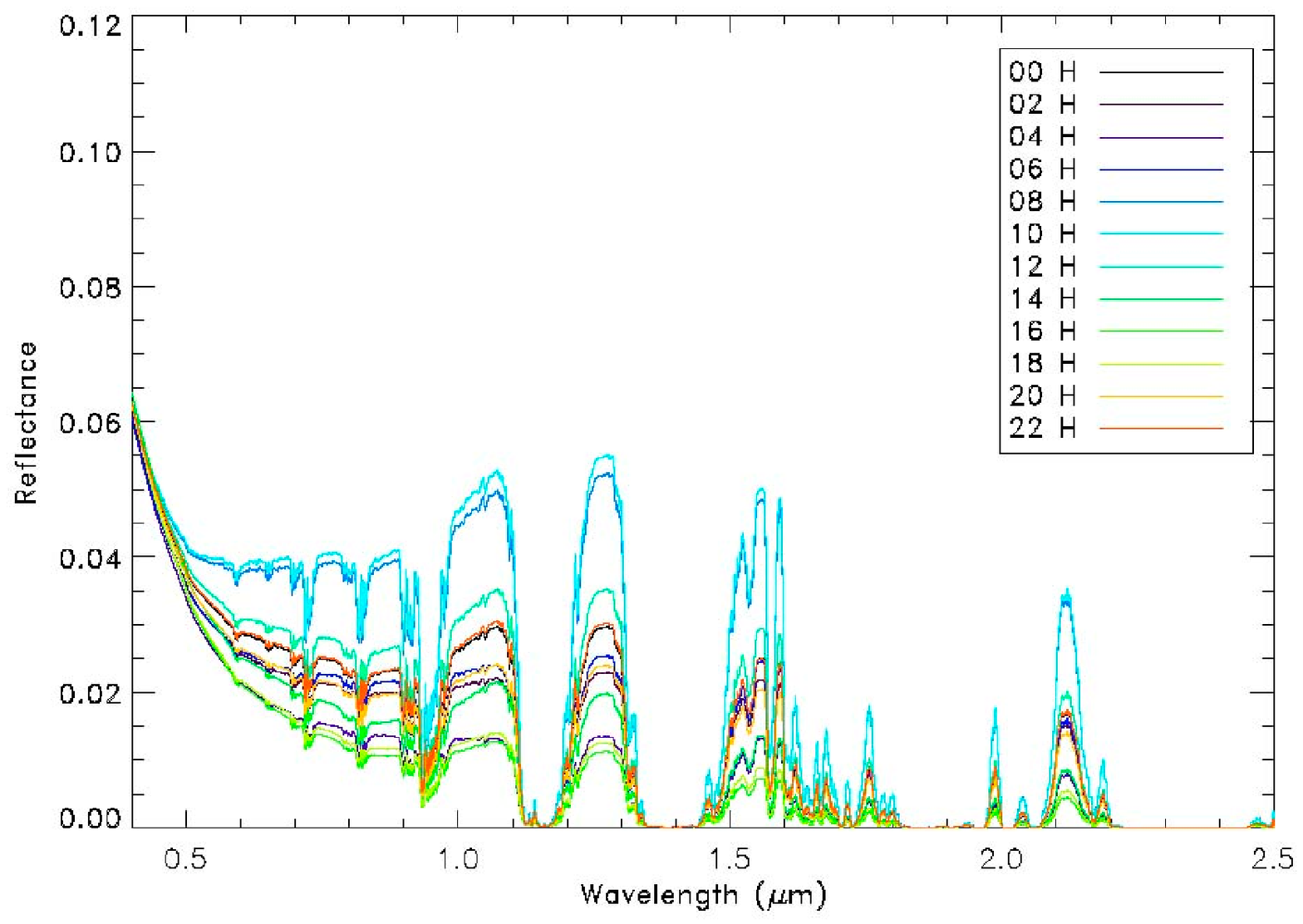} & 
   \includegraphics[width=0.49\textwidth]{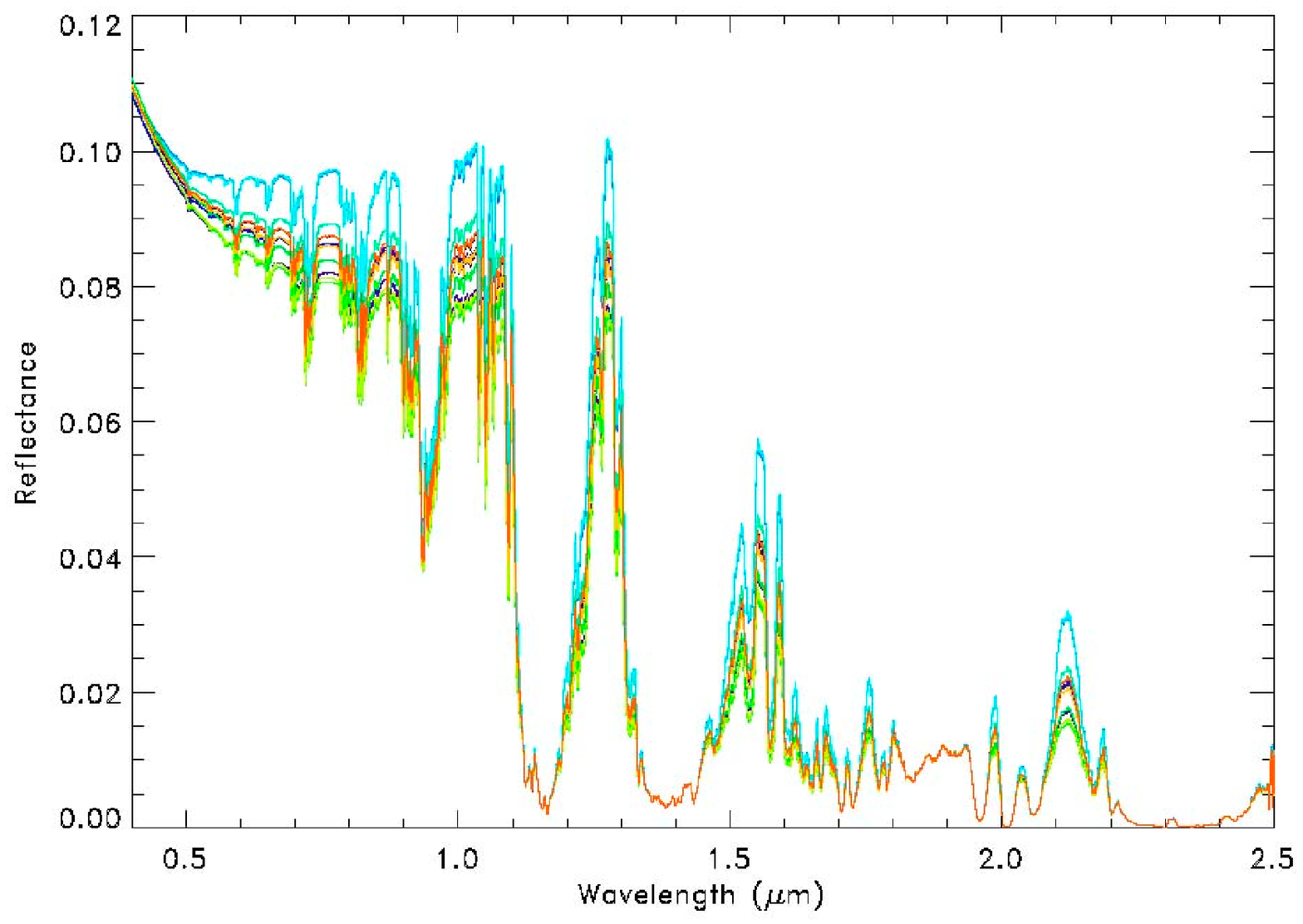}}
   \xymatrix@=0.5cm{
   \includegraphics[width=0.49\textwidth]{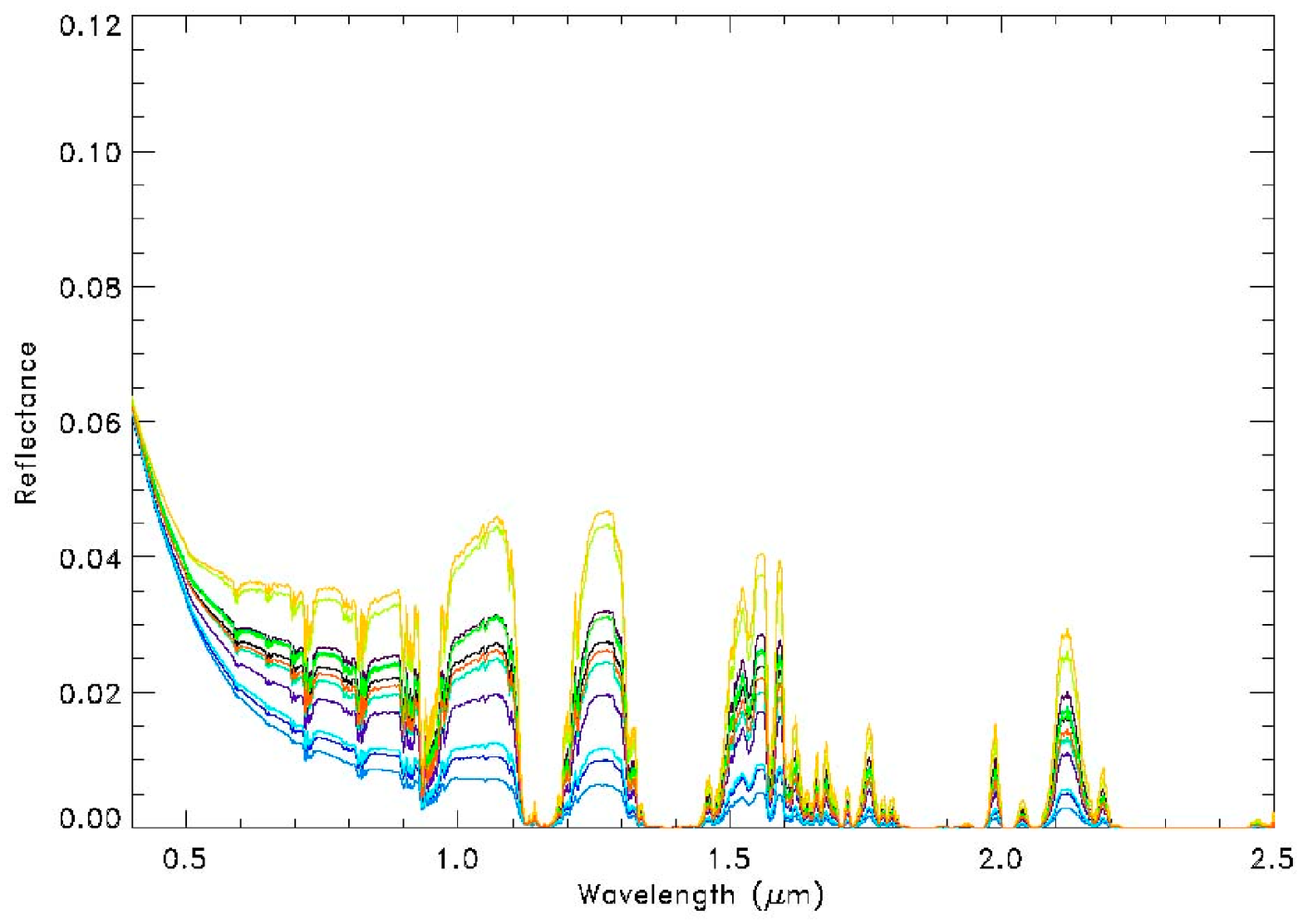} & 
   \includegraphics[width=0.49\textwidth]{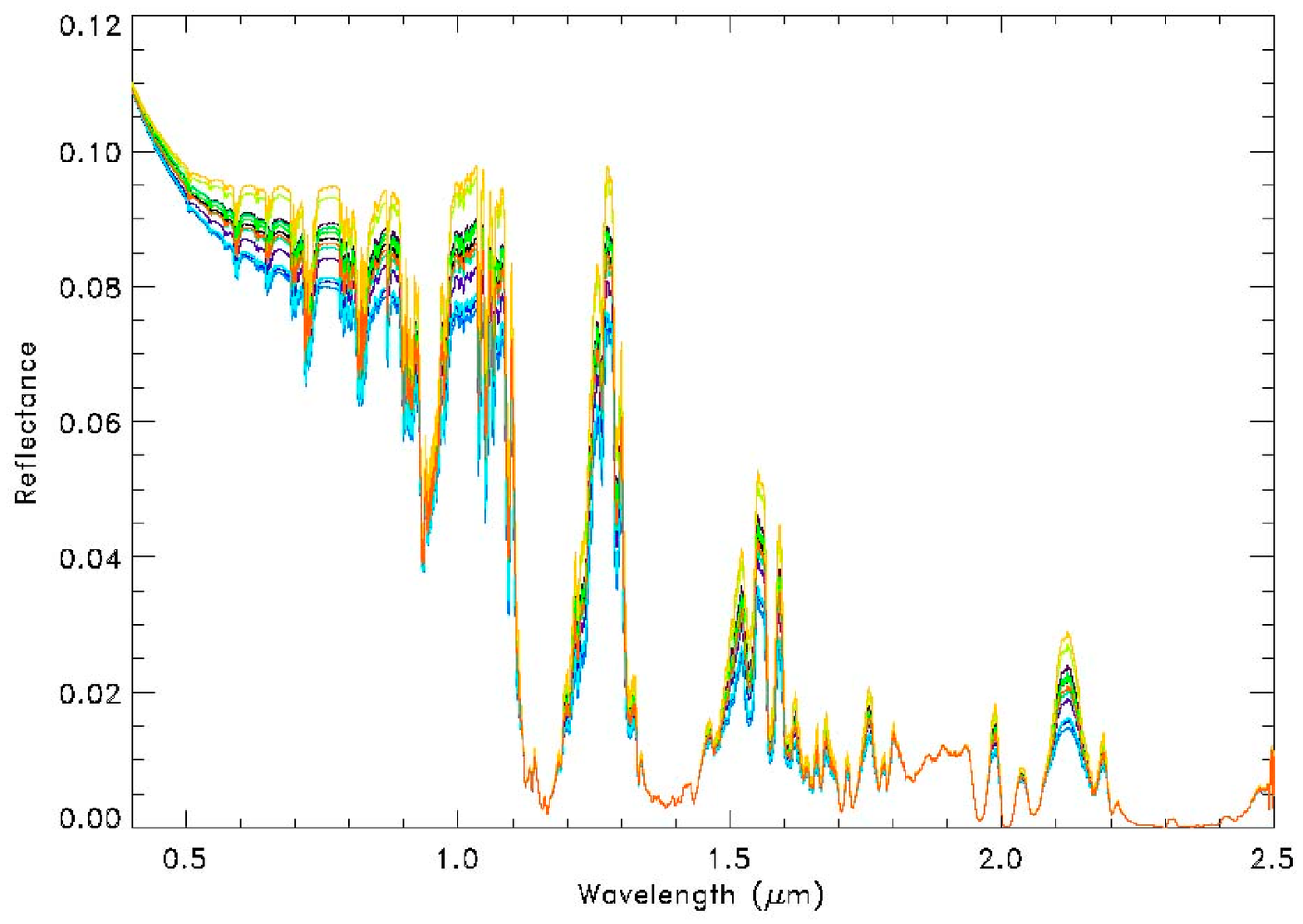}}

   \caption{Visible and near-infrared Earth's reflectance spectra 3.0 Ga ago, taken as $\pi$ times the disk-integrated radiance 
    divided by the solar flux, over the course of a day. Continents are totally desert and coastal points that are 
    closest to continent are totally populated by purple bacteria, and coastal points that are closest to oceans 
    are a mixture of purple bacteria, $10\%$, and water, $90\%$. Left panels show a cloud-free atmosphere and 
    the right panels show a cloudy atmosphere. 
    We have assumed that clouds cover $50\%$ of the surface and continental distribution corresponds to that of the 
    Earth 500Ma ago (top panels) and that of present Earth (bottom panels). The spectra have been smoothed with a 100 point running mean for display purposes.}
    \label{fig.500Ma_100_10_nubes}%
    \end{center}
    \end{figure*}
    

   \begin{figure*}
   \centering
      \xymatrix@=0.5cm{
   \includegraphics[width=0.49\textwidth]{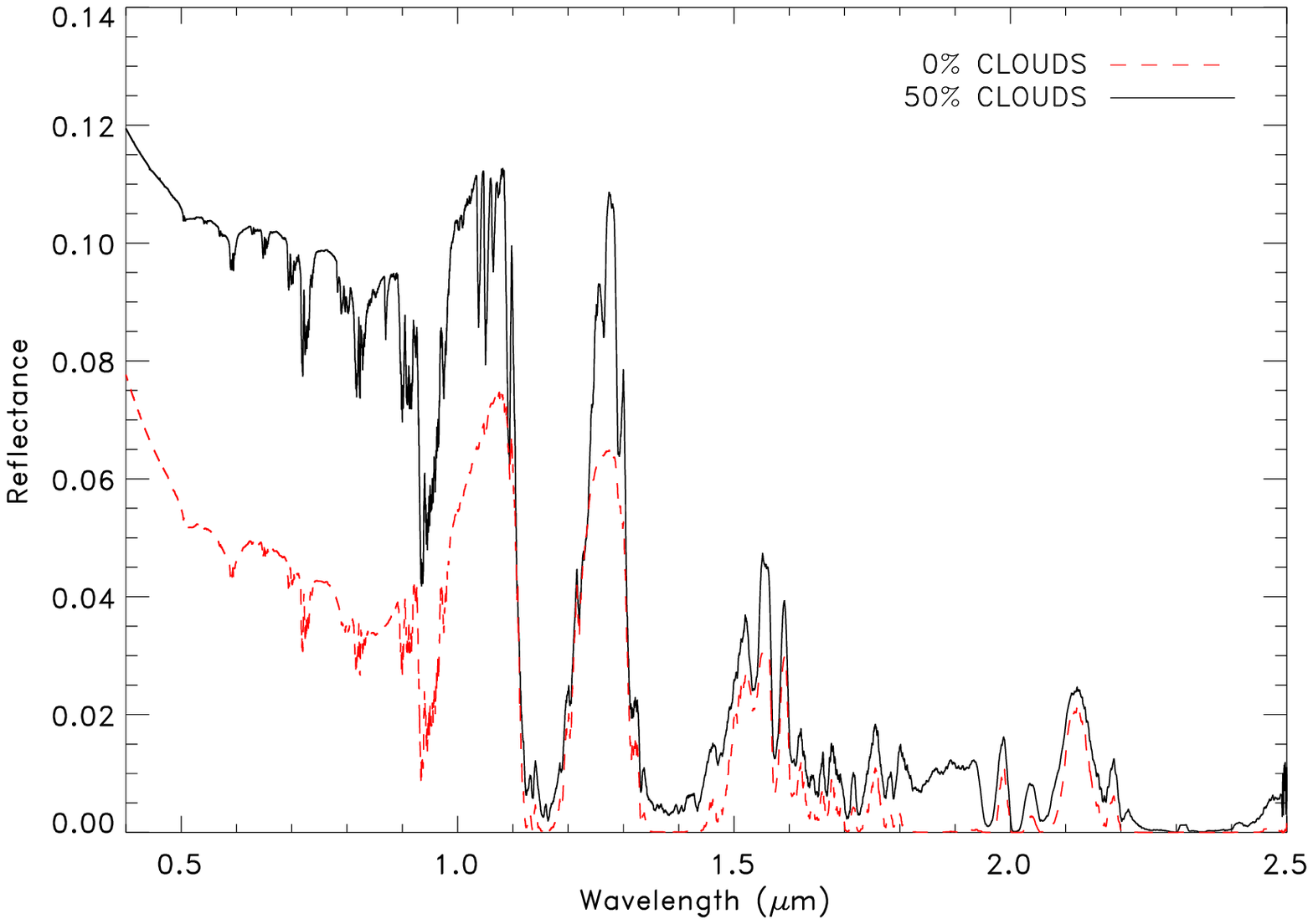}&
   \includegraphics[width=0.49\textwidth]{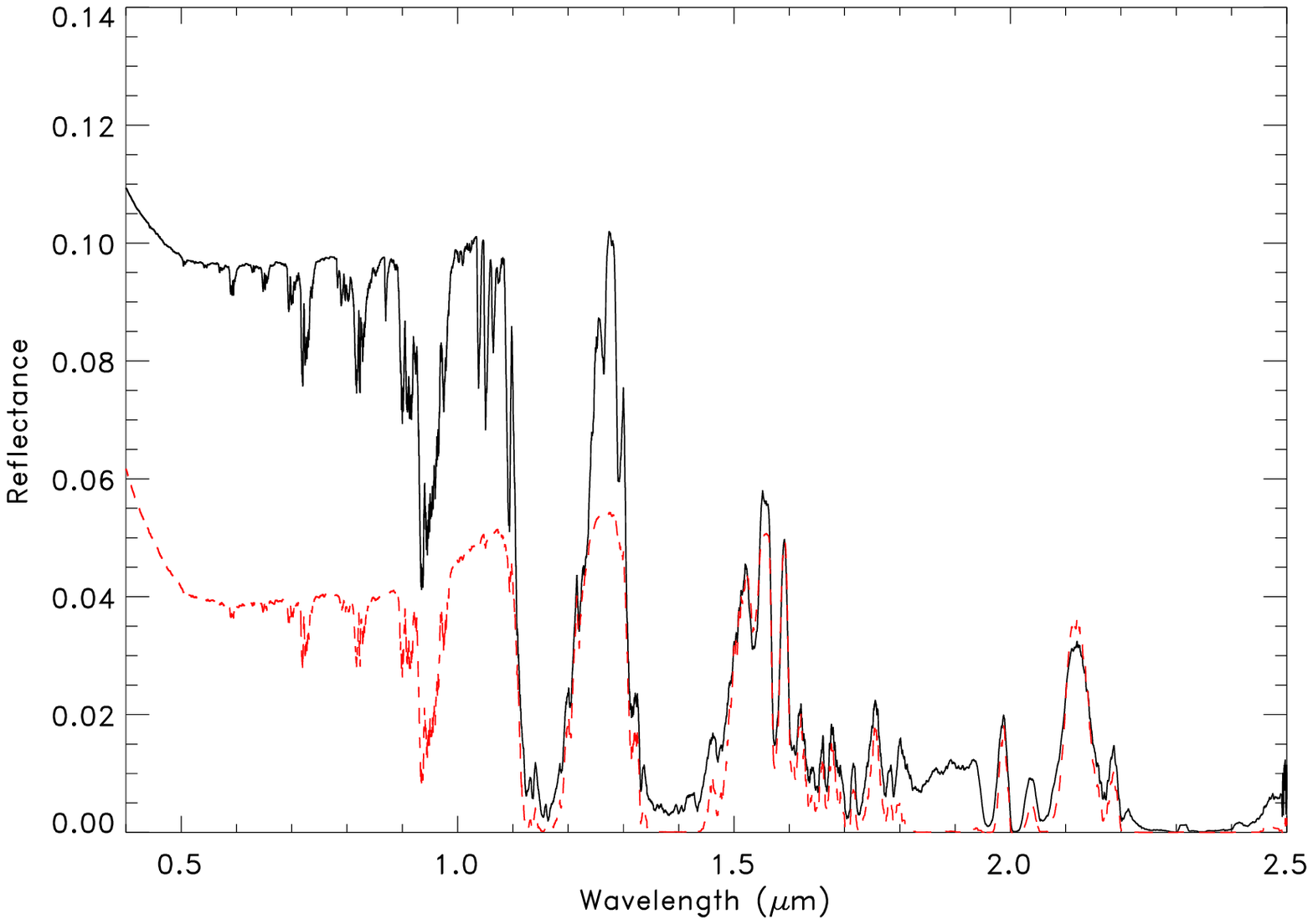}}
   \caption{Visible and near-infrared disk-integrated spectra of the early Earth for a cloud-free 
atmosphere (red dashed lines) and a 50\% cloudy 
   atmosphere (black solid lines). In the left panel, continents are totally covered by mats of 
purple bacteria and oceans are a mixture of water and purple bacteria ($90\%$ and $10\%$, respectively). 
The same is shown in the right figure but 
   here continents are bare deserts.} 
              \label{fig.bac_10_noCL_conCL}%
    \end{figure*}

   \begin{figure*}
   \centering
      \xymatrix@=0.5cm{
   \includegraphics[width=0.49\textwidth]{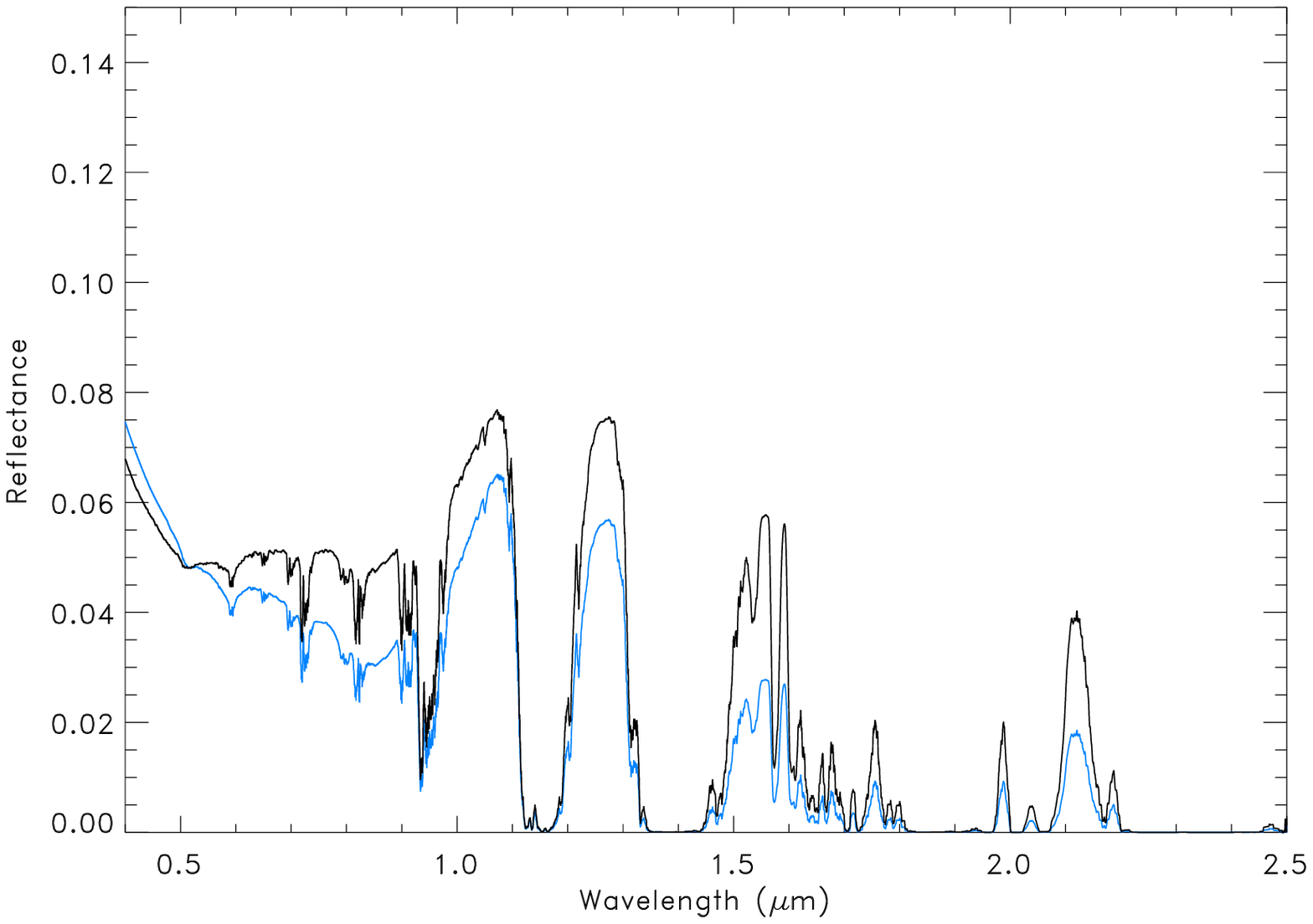} &
   \includegraphics[width=0.49\textwidth]{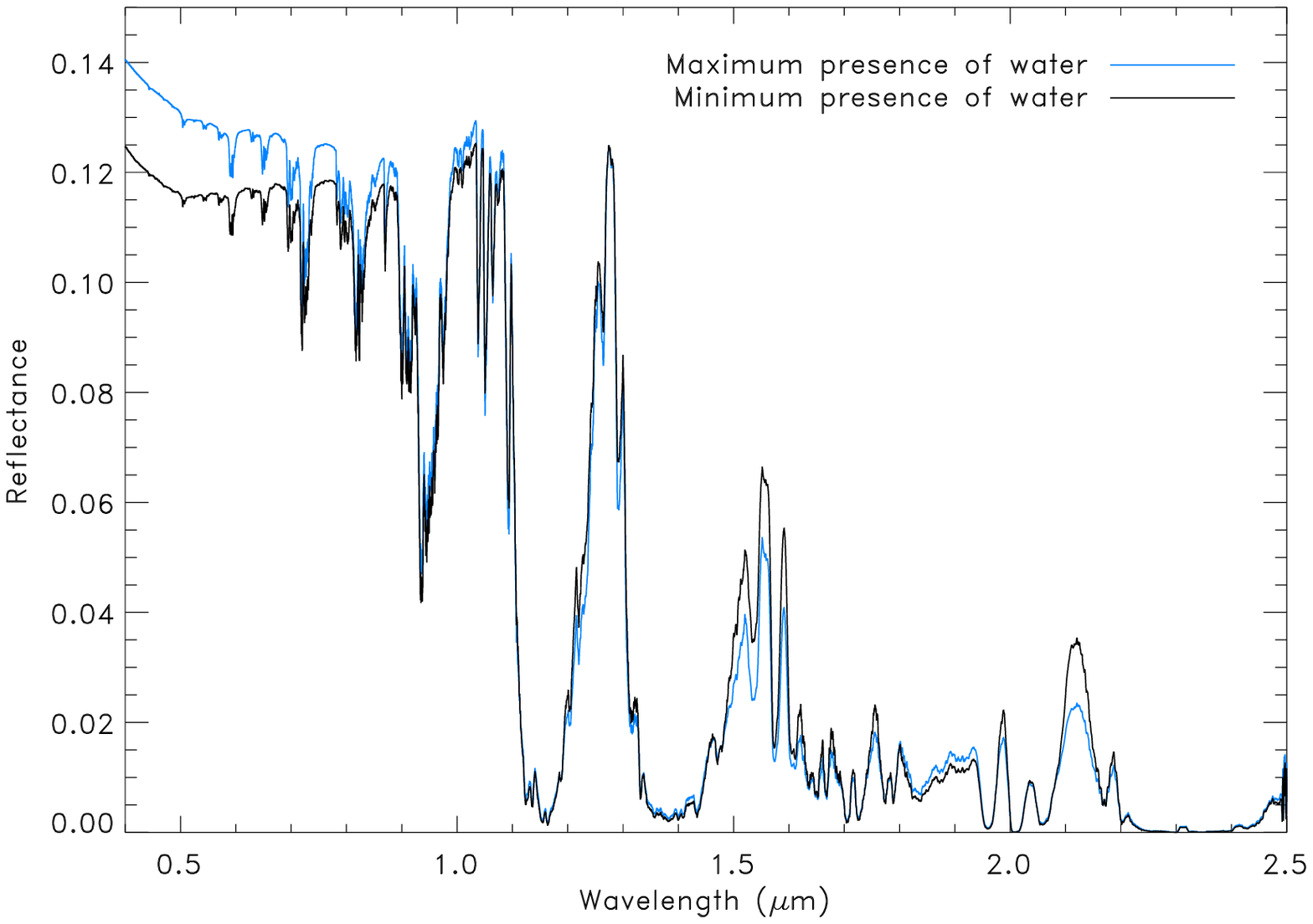}}

   \caption{Disk-averaged spectra of Earth for cloud-free (left) and a cloudy atmosphere (right). 
Here the continental and cloud distribution is that of present-day Earth and the atmospheric 
composition corresponds to that of the early Earth (3.0 Ga ago). Continents are assumed to be deserts, 
coastal zones are completely populated with purple bacteria, and oceans are a mixture of water and 
purple bacteria according to the present-day chlorophyll a distribution. Blue lines represent when 
oceans dominate de field of view and black lines when continents do.} 
                                                                                             
              \label{fig.bac_chl}%
    \end{figure*}


   \begin{figure*}
   \begin{center}
   \includegraphics[width=5.3in]{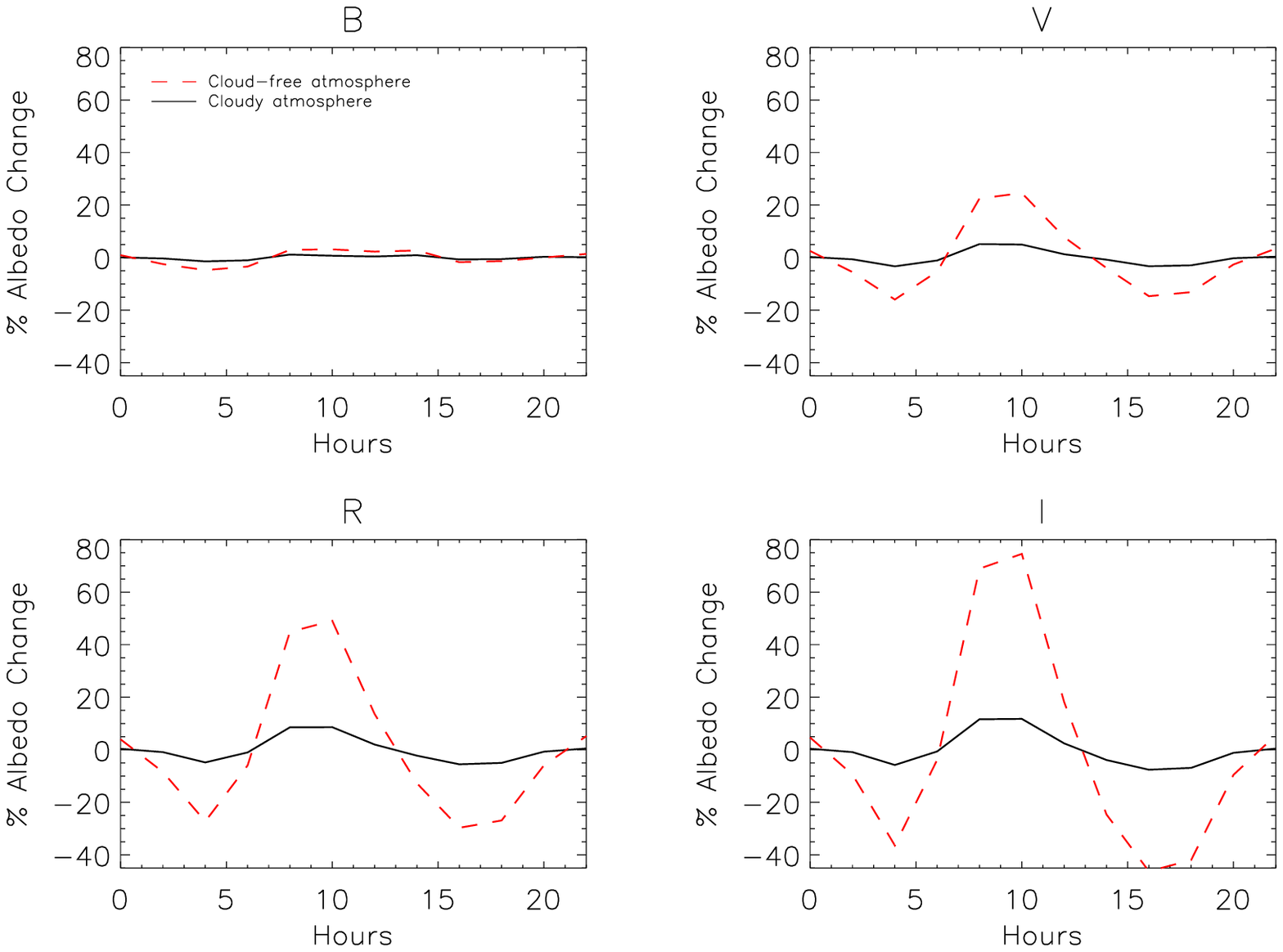}\\
   \includegraphics[width=5.3in]{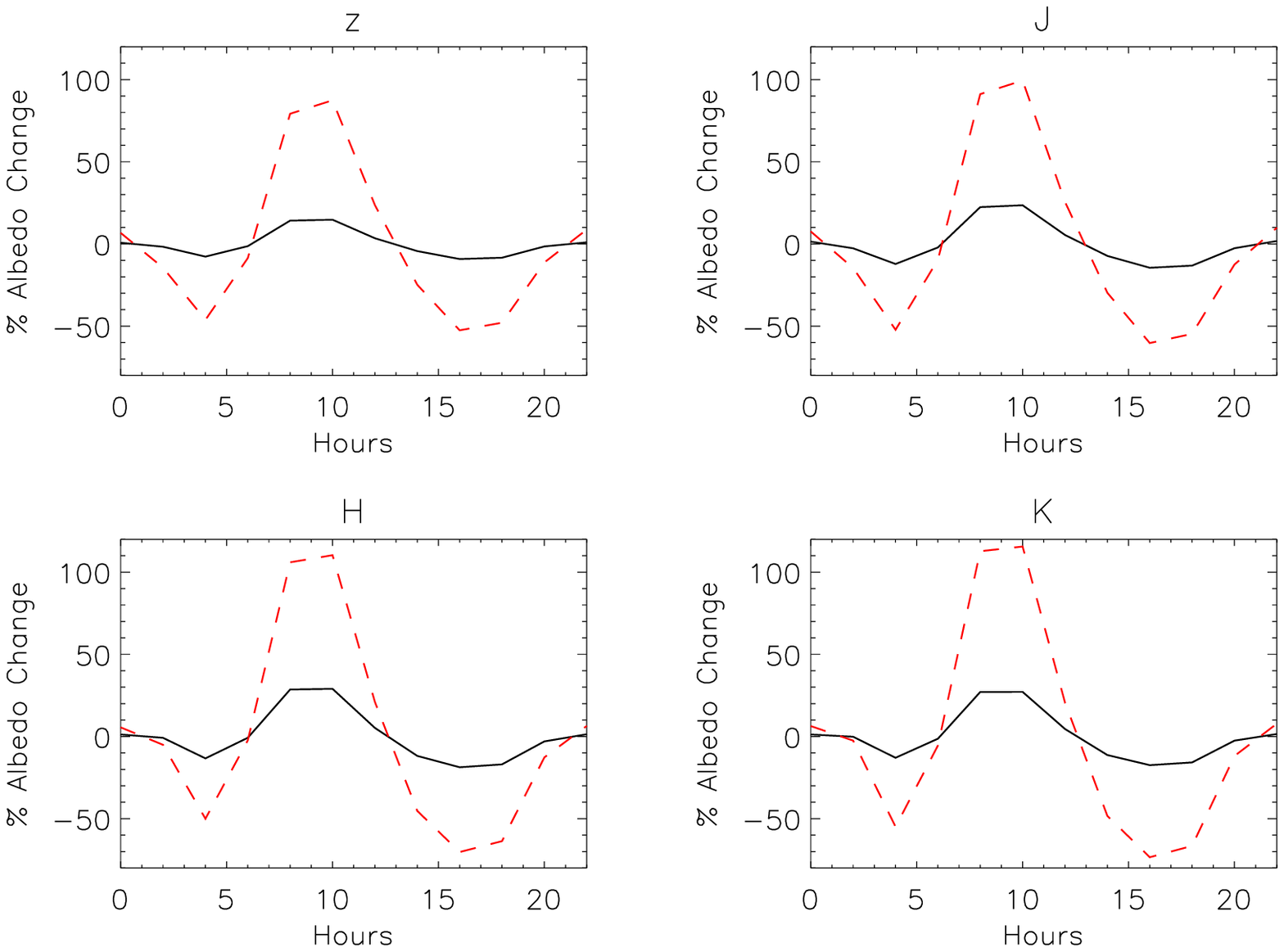}
   \caption{Daily variations of the light reflected by the early Earth for a cloud-free (red lines) and for a 
   cloudy atmosphere (black lines). Cloud cover is assumed to be 50\%. The continental distribution is that of the Earth 500Ma ago. Continents are completely covered by deserts, coastal land areas are covered by purple bacteria mats, and oceanic coastal areas are a mixture of 10\% bacteria and 90\% water. }
              \label{UBVRI_500D_0cl}%
    \end{center}
    \end{figure*}
    

   \begin{figure*}
   \centering
   \includegraphics[width=5.3in]{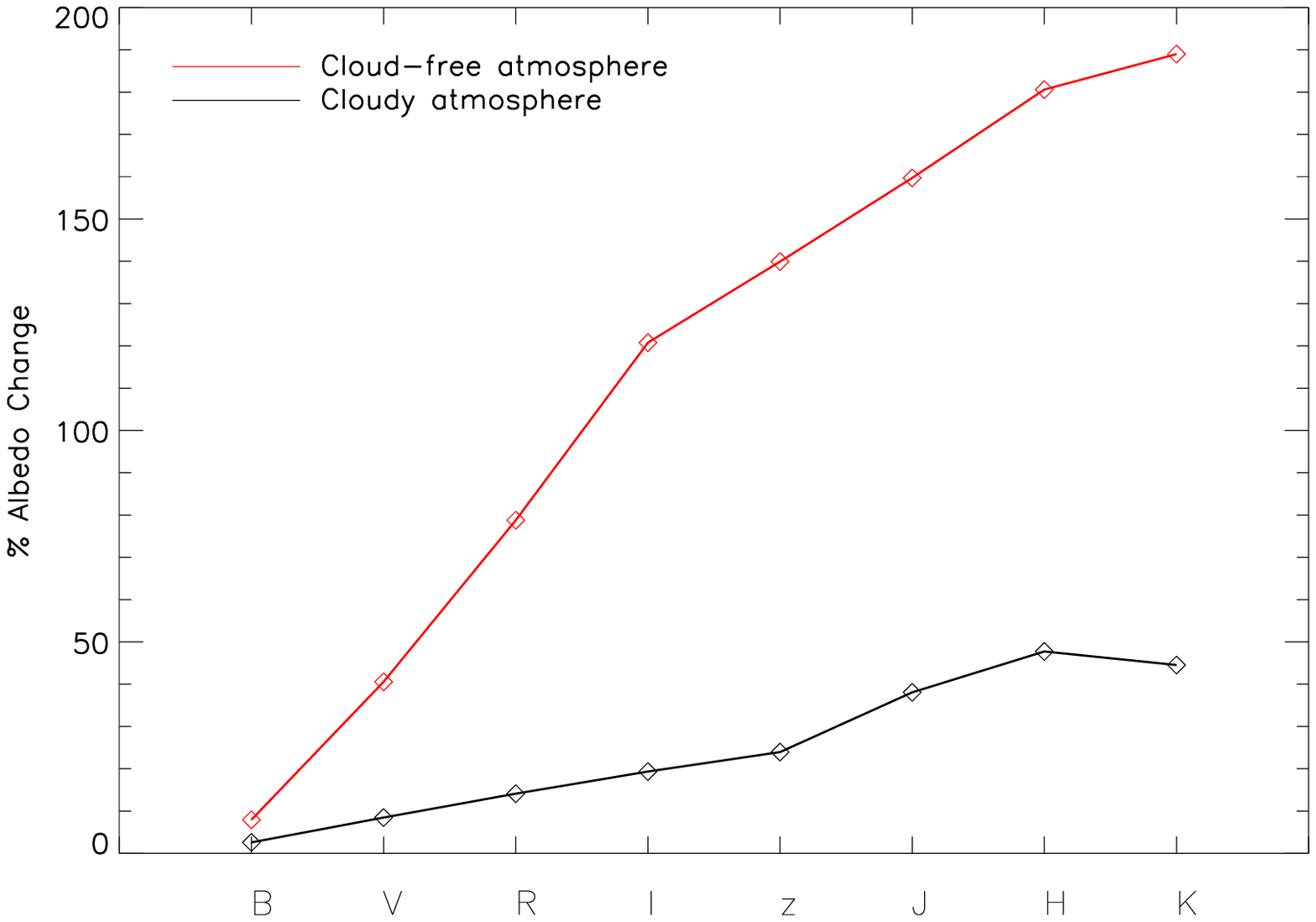}\\
   \includegraphics[width=5.3in]{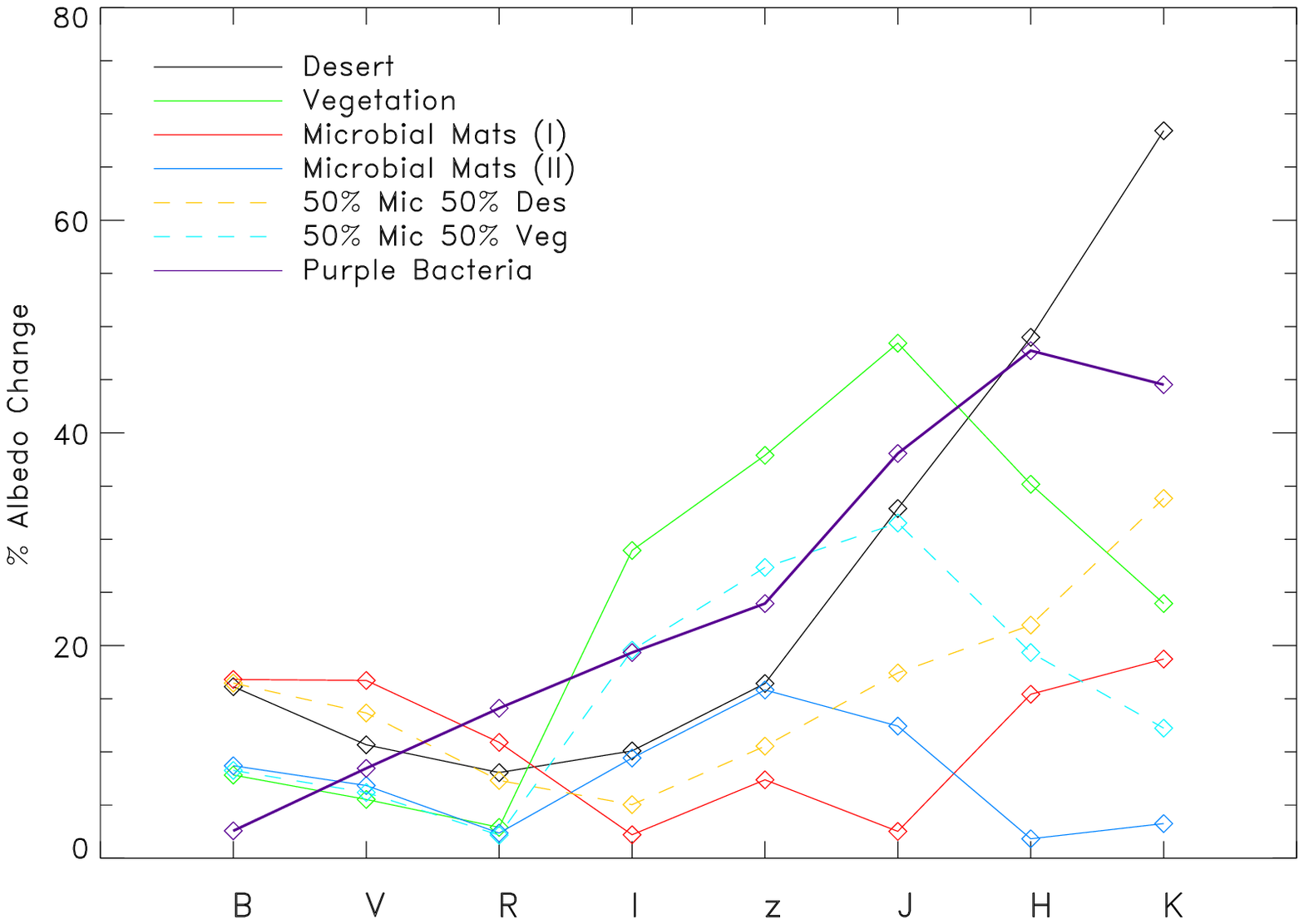}

   \caption{Top: Amplitude of the albedo variability of the early Earth as a function of the standard photometric filters. 
   Continents are covered by deserts and coasts are a mixture of purple bacteria and water. Black represents a cloudy 
   atmosphere and red a cloud-free atmosphere. Bottom: Figure 6 published in \citet{San13} including the amplitude 
   variability of the cloudy case shown in the top of this Figure.}
    \label{fig.BVRI_zJHK}%
    \end{figure*}

\begin{table}
\begin{center}
\caption{Strength of the purple bacteria signal as a function of bacteria percentage in oceans.
\label{tbl-1}}
\begin{tabular}{ccc}
\tableline
Bact. in Water & No Clouds & With Clouds\\
\tableline
10\%  						&0.0295 &0.0088\\
20\%   	 					&0.0400 &0.0140\\
30\%			 			&0.0505 &0.0193\\
40\% 						&0.0610 &0.0245\\
50\% 						&0.0715 &0.0298\\
\tableline
\end{tabular}

%


\tablecomments{Slope between the intensity in the 0.745-0.770 $\mu$$m$ and the 1.010-1.034 $\mu$$m$ range.
}
\end{center}
\end{table}

\end{document}